%% file: colm2025_conference.tex
\makeatletter
\g@addto@macro\bfseries{\boldmath}
\makeatother

\documentclass{article} 
\usepackage[final]{colm2025_conference}

\usepackage{microtype}
\usepackage{hyperref}
\usepackage{url}
\usepackage{booktabs}
\usepackage{wrapfig}

\usepackage{lineno}

\usepackage{graphicx}
\usepackage{subfigure}
\usepackage{colortbl}
\usepackage{inconsolata}
\usepackage{nicematrix}
\usepackage{enumitem}
\usepackage{multirow, array}

\usepackage{algorithm}
\usepackage{algpseudocodex}

\newcommand{\method}[0]{\textsc{UTGen}}
\newcommand{\debugmethod}[0]{\textsc{UTDebug}}

\definecolor{comm}{gray}{0.5}
\usepackage{enumerate}

\usepackage{amsmath}
\usepackage{amssymb}
\usepackage{mathtools}
\usepackage{amsthm}

\definecolor{darkblue}{rgb}{0, 0, 0.5}
\hypersetup{colorlinks=true, citecolor=darkblue, linkcolor=darkblue, urlcolor=darkblue}

\usepackage[capitalize, nameinlink]{cleveref}
\crefname{section}{Sec.}{Secs.}

\usepackage{listings, bbm, xcolor}
\definecolor{our_green}{rgb}{0.0, 0.5, 0.0}
\newcommand{\sfname}[1]{$\tt{\textcolor{our_green}{#1}}$}
\newcommand{\edited}[1]{\textcolor{black}{#1}}
\newcommand{\cameraready}[1]{\textcolor{black}{#1}}

\usepackage[framemethod=TikZ]{mdframed}
\definecolor{OurColor}{HTML}{36aa70}
\definecolor{UserExampleBg}{HTML}{ffffff}
\definecolor{UserExampleTitle}{HTML}{545f7f}
\newmdenv[
    roundcorner=5pt,
    backgroundcolor=UserExampleBg,
    linecolor=UserExampleTitle,
    outerlinewidth=0.5pt,
    frametitlebackgroundcolor=UserExampleTitle,
    frametitlefont={\bfseries\color{white}},
]{user_example}

\AtBeginEnvironment{user_example}{\setlength{\parindent}{0pt}}

\title{Learning to Generate Unit Tests for Automated Debugging}

\author{\textbf{Archiki Prasad}\quad\quad
    \textbf{Elias Stengel-Eskin} \quad\quad
    \textbf{Justin Chih-Yao Chen} \\
    \textbf{Zaid Khan} \qquad\qquad
    \textbf{Mohit Bansal} \vspace{5pt} \\
    University of North Carolina at Chapel Hill \\
\texttt{\{archiki, esteng, cychen, zaidkhan, mbansal\}@cs.unc.edu}
}

\begin{document}

\ifcolmsubmission
\linenumbers
\fi

\maketitle
\begin{abstract}
Unit tests (UTs) play an instrumental role in assessing code correctness 
as well as providing feedback to large language models (LLMs),
 motivating automated test generation. 
However, we uncover a trade-off between generating unit test inputs that \textit{reveal errors} when given a faulty code and \emph{correctly} predicting the unit test output without access to the gold solution. 
To address this trade-off, we propose \method{}, which teaches LLMs to generate unit test inputs that reveal errors along with their correct expected outputs based on task descriptions.
Since model-generated tests can provide noisy signals (e.g., from incorrectly predicted outputs), we propose \debugmethod{} that (i) scales \method{} via test-time compute to improve UT output prediction, and (ii) validates and backtracks edits based on multiple generated UTs to avoid overfitting, and helps LLMs debug effectively.
\edited{We show that \method{} outperforms other LLM-based baselines by 7.59\% based on a metric measuring the presence of \emph{both} error-revealing UT inputs and correct UT outputs. When used with \debugmethod{}, we find that feedback from \method{}'s unit tests improves pass@1 accuracy of Qwen2.5 32B on HumanEvalFix and our own harder debugging split of MBPP+ by over 3.17\% and 12.35\% (respectively) over other LLM-based UT generation baselines. } \cameraready{Moreover, we observe that feedback from Qwen2.5 32B-based \method{} model can enhance debugging with frontier LLMs like GPT-4o by 13.8\%.}
Lastly, we demonstrate that \method{} is a better judge for code correctness, outperforming a state-of-the-art trained 8B reward model by 4.43\% on HumanEval+ with best-of-10 sampling using Qwen2.5 7B. Our code and datasets are publicly available at \url{https://github.com/archiki/UTGenDebug}.
\end{abstract}

\vspace{-0.25em}
\section{Introduction}

\begin{wrapfigure}{r}{0.575\textwidth}
    \vspace{-2.25em}
    \centering
    \includegraphics[width=0.9\linewidth]{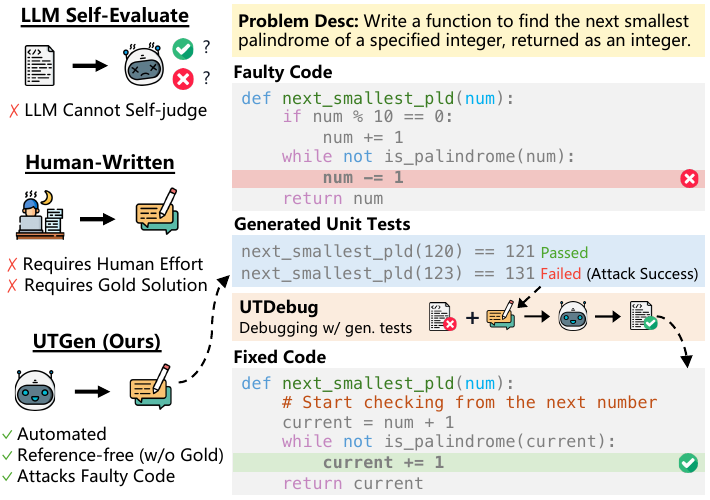}
    \vspace{-1.25em}
    \caption{
    We propose \method{}, which automatically generates failing unit tests (UTs) for a faulty code (triggering errors) without access to the gold solution. The generated UTs can in turn be used for LLM debugging via \debugmethod{}, improving code accuracy.}
    \label{fig:fig1}
    \vspace{-1em}
\end{wrapfigure}
With rapid advancements in training large language models ~\citep[LLMs;][]{achiam2023gpt,claude3modelcard,team2023gemini}, enhancing their coding abilities has garnered significant attention~\citep[\emph{inter alia}]{chen2021evaluating,li2022competition,li2023starcoder, roziere2023code,guo2024deepseek}. 
However, these models are far from perfect and -- much like human-written code -- model-written code contains errors.
Human developers often improve their code through test-driven development, i.e., identifying failure cases by providing example inputs and their expected outputs -- referred to as \textit{unit tests} (UTs) -- to test individual functionalities, reasoning over causes of failure, and modifying the code to address the issues 
~\citep{tdd_at_ibm,quality_improvement_tdd, ficco2011bug,beck2022test}. 
Models have similarly been shown to benefit from iterative debugging based on explicit or implicit feedback stemming from failing unit tests~\citep{zhang2023self,chen2023teaching,moon2023coffee}.
However, the ability to provide feedback and debug incorrect code is often bottlenecked by the availability of (failing) unit tests for a given problem.

While several coding benchmarks come with 
human-written UTs for evaluation purposes, these suites of UTs are typically small due to the laborious annotation process~\citep{liu2024your}. 
Unit test collection is challenging for two reasons: first, it requires sampling inputs that are likely to trigger an error. For example in \cref{fig:fig1}, the unit test: {\sfname{next\_smallest\_pld}\tt(120)==121} would \emph{not} trigger any error (despite the fact that the code is incorrect for non-multiples of 10), 
while another unit test: {\sfname{next\_smallest\_pld}\tt(123)==131} would lead to an incorrect output, revealing a bug in the function.
Secondly, UTs require expected outputs (e.g., {\tt 131} in the previous UT), i.e., the desired behavior of the code being tested must be known.
Due to these challenges, prior work employs LLMs to generate unit test inputs at random (i.e., without conditioning on faulty code to generate failing inputs) and often uses the gold (correct) code solution for generating outputs~\citep{chencodet,chae2024coffee}.
Therefore, these approaches do not scale well at inference time, which requires an \emph{online and automated} method for generating UTs as the LLMs solve problems on the fly. To bridge this gap, we pose the following research questions:

\begin{itemize}[topsep=1pt, noitemsep,leftmargin=*]
    \item \textbf{RQ1:} \emph{What are desirable properties for UT generators, and how do we measure them?}
    
    \item \textbf{RQ2:} \emph{How well do LLMs perform zero-shot UT generation, and how can we improve them?}

    \item  \textbf{RQ3:} \emph{How do we best use an automated but potentially noisy UT generator to improve code debugging and generation?}
\end{itemize}

To address RQ1, we characterize two desirable properties of unit test generators (in \cref{ssec:setup}): 
 1) \emph{high attack rate}, i.e., given a faulty code, the unit test generator should generate inputs that are likely to trigger errors;\footnote{Here, an `error' means either a syntax/runtime error, or when the result of the target code does not match the expected output of the UT. We expand on this in \cref{sec:method}.} 2) \emph{high output accuracy}, ensuring that the UT output is consistent with the task description (and that of a correct solution). 
For instance, in \cref{fig:fig1}, \sfname{next\_smallest\_pld}{\tt(120)} would lead to a lower attack rate, as it does not trigger any errors, while  \sfname{next\_smallest\_pld}{\tt(123)==131} \textit{does} (high attack rate); however, both have high output accuracy as in both cases the output is correct. We measure these properties via three \emph{intrinsic metrics}: measuring attack rate and output accuracy independently and then, crucially, how often a UT generator can generate \textit{both} failing inputs along with correct UT outputs based only on the problem description and a faulty code.
We benchmark the ability of LLMs to act as zero-shot unit test generators and show that coding LLMs generally struggle (cf. \cref{ssec:intrinsic}). 

Moreover, addressing RQ2, we find that, without finetuning, zero-shot models often exhibit a strong trade-off between attack rate and output accuracy. In other words, the UTs that are most likely to trigger errors (i.e., higher attack rate) are generally more challenging edge cases, making it harder for the model to predict their output (i.e., lower output accuracy). 
Due to a lack of dedicated datasets for unit test generation in prior work,  we introduce \method{}, a data creation and training recipe (\cref{ssec:training}) that bootstraps training data for UT generation from \emph{existing} code generation datasets by perturbing code to simulate errors, generating failing unit test and augmenting it with chain-of-thought rationales~\citep{wei2022chain}. \edited{We show that training via \method{} better balances this trade-off and yields a higher number of unit tests that have \emph{both} attacking inputs and correct outputs, with 7.59\% absolute improvement over prompting models to generate failing UTs.}

Finally, we examine RQ3, utilizing our generated UTs on downstream, i.e., \emph{extrinsic} tasks: code debugging and code generation. 
Here, automated UT generation methods face additional challenges: unlike human-generated gold UTs -- which have correct outputs but require human involvement -- generated UTs provide noisy feedback, as the UT might fail to reveal the buggy code's errors or have an incorrectly predicted output.
To mitigate this issue, we propose \debugmethod{}, an improved multi-turn debugging method that addresses these challenges in two ways (cf. \cref{ssec:debug}): (i) we improve the output accuracy of generated UTs by scaling test-time compute via self-consistency \citep{wang2022self}; (ii) we regularize the debugging process by generating multiple UTs and accepting code edits \emph{only if} the revised code passes \emph{more} generated UTs, backtracking edits otherwise.
We then plug UTs generated from multiple LLM-based methods into \debugmethod{} in \cref{sec:extrinsic}, finding that on our most challenging subset of MBPP+ problems, 
\edited{\debugmethod{} with UTs generated by \method{} improves pass@1 accuracy of Qwen2.5 Coder 32B Instruct by 15.07\% (absolute) compared to debugging without UTs, and by 4.61\% over a zero-shot unit test generation baseline (with similar trends on 8B-scale LLMs).}
We demonstrate that by generating UTs with \method{} and computing the rate of passing UTs, we can \emph{better judge code correctness} than using scores from trained, state-of-the-art 8B reward model (RM). For best-of-10 sampling with Qwen2.5 Coder 7B Instruct, \method{} improves accuracy by 4.43\% over the RM on HumanEval+. \cameraready{Finally, in \cref{ssec:frontier}, we highlight the importance of high-quality unit tests generated by \method{} when debugging with a frontier LLM like GPT-4o~\citep{hurst2024gpt}, finding that using \method{}-feedback from a smaller model, such as Qwen2.5 32B, significantly outperforms self-generated feedback from GPT-4o by nearly 25\% (absolute) on the most challenging subset of MBPP+ problems.}

\vspace{-0.25em}
\section{Related Work}
\vspace{-0.25em}
\label{sec:rel}

\noindent\textbf{Automatic Unit Test Generation. \hspace{0.5em}} Manually writing unit tests is laborious and often infeasible~\citep{chencodet,liu2024your}. Consequently, past research explores automatic UT generation~\citep[][\emph{inter alia}]{king1976symbolic,cadar2008klee,holler2012fuzzing,cha2015program}. The advent of LLMs has spurred recent efforts in using them for UT generation~\citep{chencodet,schafer2023empirical,liu2024your}. Specifically, \citet{schafer2023empirical} and \citet{liu2024your} focus on generating \emph{unit test inputs} via prompting frontier LLMs like GPT-4 and/or iterative prompting, assuming access to the \emph{gold} solution. 
In contrast, our models, trained with \method{}, generate \emph{both input-output UT pairs} based on the task description without relying on the gold implementation. 
While \citet{chencodet} also generate input-output UT pairs using standard LLM prompting, their primary focus is code generation -- \emph{not} the quality of generated UTs. In contrast, we directly model the desiderata (or quality) of UTs, and demonstrate its utility on code generation and debugging.

\vspace{0.35em}
\noindent\textbf{LLM Debugging. \hspace{0.5em}} Using LLMs for debugging faulty code, or program repair, has been extensively studied. Debugging approaches are divided into those training models to debug~\citep{moon2023coffee,ninext,chae2024coffee} and those providing external feedback to pretrained models~\citep{chen2023teaching,zhang2023self,olausson2023self,zhong2024ldb}. Both rely on \textit{gold} unit tests for training or feedback. Thus, \method{} complements both methods by providing generated unit tests when human-written tests are scarce or unavailable.  In \cref{ssec:debug}, we introduce \debugmethod{}, a debugging pipeline that addresses noisy feedback from inaccurate unit tests through test-time scaling and backtracking. Moreover, in \cref{sec:results} we show that \method{}'s unit tests can effectively provide feedback to LLMs for code generation and debugging.

\vspace{-0.25em}
\section{Unit Test Generation and Automated Debugging}
\label{sec:method}
\vspace{-0.5em}
\subsection{Background and Task Setup}
\label{ssec:setup}
\vspace{-0.25em}

Given a natural language task description $d$ for a coding task, we focus on generating unit tests (UTs) in the format of \emph{input-output pairs} that are consistent with the task description $d$. 
Our setup is consistent with \citet{chencodet} and \citet{jain2024livecodebench} who also consider unit tests in the form of input-output pairs generated \emph{without} utilizing the correct implementation of the function. 
More generally, our setting is akin to parameterized unit testing~\citep{parameterized_unit_testing} and uses the notion of \emph{functional correctness}, i.e., measuring correctness by simulating an exhaustive set of scenarios (UT inputs) and ensuring that the function performs as expected (as per the problem description $d$).

\vspace{-0.25em}
\paragraph{Notation and Desiderata.\hspace{0.75em}} 
Let $d$ denote the natural language description of the function to be implemented (top yellow box in \cref{fig:fig1}).
We assume that this description specifies the input space $\mathcal{X}$  and output space $\mathcal{Y}$. 
Furthermore, let the set of all functions 
that correctly solve the task be $\mathcal{F}_d$. 
Then, a \emph{valid} unit test $(x,y)$ for task $d$ is: 

\begin{itemize}[nosep,leftmargin=*]
    \item $x \in \mathcal{X}$, i.e., the input of the unit test is a \emph{valid input} as per the task description $d$. 
    \item $f_r(x) = y, \forall f_r \in \mathcal{F}_d$, i.e., $y$ is the \emph{expected output} as per the task description $d$, and therefore, is the result one gets by executing any correct implementation. 
\end{itemize}

\noindent For example, in \cref{fig:fig1}, {\tt 120} is a valid input, as it is a number, whereas {\tt "apple"} is not. 
Similarly, {\tt 121} is the expected output of the function (given {\tt 120} as input), while {\tt 122} would be an invalid output. 
Thus, {\tt (120, 121)} is a valid unit test for the task. 

\paragraph{Unit Test Generator.} Addressing RQ1, we define the desirable properties of an automatic unit test generator $T_\theta$, parameterized by $\theta$.
Ideally, $T_\theta$ should generate \emph{valid unit tests} from a task description $d$, i.e., $T_\theta(d) \mapsto (x,y)$ without any manual intervention. 
However, to account for downstream utility of unit tests, we denote a potentially buggy code under testing as $\widehat{f}_b$.
If $\widehat{f}_b \notin \mathcal{F}_d$ is faulty or implemented incorrectly, 
a \emph{desirable} unit test generator should be able to efficiently generate \emph{failing unit tests}: 
$T_\theta(d, \widehat{f}_b) \mapsto (x, y)$, such that $\widehat{f}_b(x) \neq y$, i.e., a valid unit test input $x$ that uncovers the errors in $\widehat{f}_b$.
Moreover, $T_\theta$ must also predict the correct UT output $y$, i.e., the output of the correct code. 
We study generators of the form $T_\theta(d)$, which consider only the description, and debugging-style generators of the form $T_\theta(d, \widehat{f}_b)$, which also consider a buggy code solution $\widehat{f}_b$. Empirically, we find that $T_\theta(d)$ lacks sufficient context to generate effective error-revealing tests. Therefore, we focus on training UT generators of the form $T_\theta(d, \widehat{f}_b)$.

\vspace{-0.25em}
\subsection{\method{}: Training LLMs for \underline{U}nit \underline{T}est \underline{Gen}eration}
\label{ssec:training}
\vspace{-0.25em}

While prior work focuses on curating training data for improving code generation \citep{deepseek-coder,octopack}, 
there is a general lack of dedicated datasets for training the desired UT generator outlined above. 
Therefore, to improve along the lines of RQ2, we design, \method{}, a training recipe that bootstraps this data from training datasets for code generation i.e., a collection of problem descriptions and their corresponding code. 

\begin{figure*}[t]
    \centering
    \includegraphics[width=0.95\linewidth]{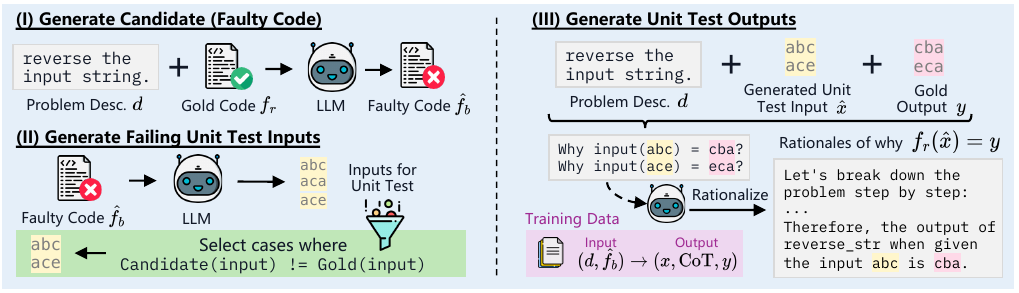}
    \vspace{-6pt}
    \caption{\textbf{\method{} Training Pipeline:} Starting with training data for code generation (problem description and gold code), we create training data for UT generation in three stages: (I) perturbing gold code to generate faulty codes, (II) generating UT inputs and filtering for failing UTs, and (III) generating and relabeling chain-of-thought rationales conditioned on the gold code's outputs.
    }
    \label{fig:method}
\end{figure*}

\vspace{-0.25em}
\paragraph{Problem Descriptions and Target Codes.} 
We start with a collection of coding problems with problem descriptions ($d$) and gold codes ($f_r$) as illustrated in \cref{fig:method} (I). 
Specifically, we use publicly available data from Tulu-3~\citep{lambert2024tulu3} due to its large scale and the improvements noted by \citet{lambert2024tulu3} when finetuning  LLMs on coding tasks.\footnote{The code solutions in the Tulu-3 SFT mix have either been written by humans or by frontier models and are thus highly likely to be correct~\citep{lambert2024tulu3}.} 
We filter it to focus on Python code with functional abstractions (further details in \cref{app:method}).
However, in order to train a unit test generator we would need access to incorrect or faulty code that can be debugged, as unit tests must have some error to attack.
We obtain these by using the LLM to perturb the reference code solution.

\vspace{-0.25em}
\paragraph{Annotating Unit Tests.} As mentioned in \cref{ssec:setup}, one of the goals of the unit test generator is to be able to not only generate valid unit tests, but \emph{failing} unit tests (that trigger errors). 
To facilitate this, given a problem description, reference code $f_r$, and buggy candidate code $\widehat{f}_b$, we sample $n$ different unit test inputs (via the prompts in \cref{app:prompts}).
A unit test $(x, f_r(x))$ is \emph{failing} if $f_r(x) \neq \widehat{f}_b(x)$, 
i.e., if the output of the candidate fails to match the reference output (cf. \cref{fig:method} (II)). 
Note that, while the LLM can be used to generate the output of unit test, it can often be inaccurate~\citep{jain2024livecodebench,gu2024cruxeval}; 
so to ensure output accuracy, we use the output of the reference code during training.
\cameraready{Due to the formalization of UT output prediction as a \emph{reasoning word problem}, we further explore generating chain-of-thought~\citep[CoT;][]{wei2022chain} reasoning before predicting the UT output. To this end, we employ the post-hoc rationalization procedure outlined in \citet{zelikman2022star} -- given the entire UT $(x, f_r(x))$, we ask the LLM to generate rationales supporting why $f_r(x)$ is the output corresponding to the input $x$. Then, to create the supervision data, we add these rationales as CoTs prior to the output prediction, as illustrated in \cref{fig:method} (III). }
We revisit the discussion of how accurate LLMs are at output prediction during inference in \cref{sec:results}.

\vspace{-0.25em}
\paragraph{Supervised Finetuning.} 
When training LLMs, the input to the LLM is the same prompt used for sampling unit tests (listed in \cref{app:prompts}) and the output is a failing unit test. We collect nearly 30K training instances for 8B-scale models and roughly 70K instances for training 32B LLMs (details in \cref{app:method}). 
The goal here is to improve both output accuracy and attack rate \emph{jointly} via supervised finetuning using the negative log-likelihood loss using basic hyperparameters outlined in \cref{app:method}.

\vspace{-0.25em}
\subsection{\debugmethod{}: Generated \underline{U}nit \underline{T}ests for \underline{Debug}ging}
\label{ssec:debug}
\vspace{-0.25em}

\begin{figure*}[t]
    \centering
    \includegraphics[width=\linewidth]{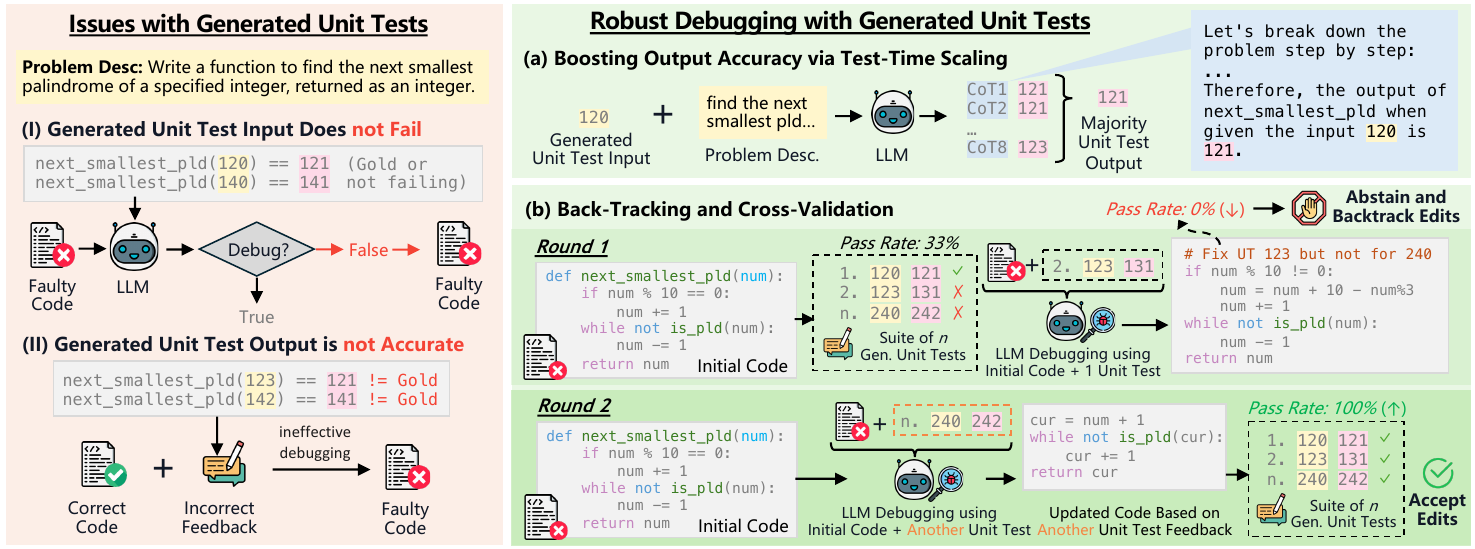}
    \vspace{-2em}
    \caption{\textbf{Left}: We highlight potential issues with debugging a faulty code using generated UTs: (I) \emph{non-failing} UTs misclassify faulty code as correct; (II) UTs with incorrect outputs produce incorrect feedback and consequently, unsuccessful debugging. \textbf{Right}: We introduce \debugmethod{} which (a) uses inference-time scaling to select better UT outputs based on a majority vote, and (b) generates multiple UTs for validation, discarding edits when overall pass rate decreases (round 1) and accepting edits when overall pass rate improves (round 2).}
    \label{fig:fig3}
    \vspace{-1em}
\end{figure*}

A key difference when using generated unit tests, as opposed to human-generated UTs, is the \emph{degree of noise} in the feedback. Despite training models via \method{} or other methods, a generated unit test $(\widehat{x}, \widehat{y})$ may not be 100\% accurate. This can manifest in two ways: 1) the generated UT \emph{input does not fail} for the code under debugging $\widehat{f}_b$, i.e., $f_r(\widehat{x}) = \widehat{f}_b(\widehat{x})$; 2) the generated UT \emph{output is inaccurate}, i.e., not consistent with what a gold solution would yield ($\widehat{y} \neq f_r(\widehat{x})$). Both types of errors can negatively impact the utility of unit tests for debugging, as shown in \cref{fig:fig3}~(left). A non-adversarial input might result in faulty code being misclassified as correct and prematurely removed from debugging. Additionally, incorrect outputs can cause false positives, introducing errors to otherwise correct code. Even if the candidate code has bugs, incorrect UT outputs can lead to incorrect feedback, thereby, degrading performance. These issues motivate our RQ3 to incorporate noisy feedback. We propose \debugmethod{}, which includes two ways to mitigate noisy feedback from automatically generated unit tests (additional details in \cref{app:debug}). We empirically validate the importance of both these components of \debugmethod{} in \cref{ssec:ablate}.

\vspace{-0.5em}
\paragraph{Boosting Output Accuracy via Test-Time Scaling.}
Building on past work that has shown the benefits of scaling up inference-time compute~\citep{wang2022self,lightman2023let,snell2024scaling}, we improve UT output accuracy by allocating additional computation to the problem.
Specifically, we use self-consistency \citep[SC;][]{wang2022self}, whereby, for a given UT input, we sample $k=8$ output completions (including CoT rationales) and take the most common final UT output (majority vote) as the final answer, as shown in \cref{fig:fig3}~(top-right). 
To further boost output accuracy, consistent with \citet{prasad2024self}, we upsample UT inputs and only retain those where the final answer gets over 50\% of the votes (i.e., 4 votes), discarding the unit test otherwise. 

\vspace{-0.5em}
\paragraph{Back-Tracking and Validation.} To handle noisy feedback, it is crucial to know when to \emph{abstain or backtrack}, e.g., discard edits. This is useful when generated UT outputs are incorrect, leading to faulty feedback, or when correct feedback is not incorporated by the model. Inspired by test-driven development, we accept changes only if the revised code \emph{passes the previously failing UT}. Moreover, developers often use multiple UTs to detect errors from changes in other parts of the code. 
This helps prevent \emph{overfitting}, where code is modified to pass a single unit test but fails others. Therefore, in each debugging round via \method{}, we generate $n$ UTs. We use one for debugging feedback and accept edits only if the pass rate on the entire set improves (validation); otherwise, we backtrack. This process is shown in \cref{fig:fig3}~(b), where edits in round 1 are discarded, but those in round 2 improve the test suite as a whole and are accepted.

\vspace{-0.25em}
\section{Experimental Setup }
\label{sec:setup}
\vspace{-0.25em}

\paragraph{Models.} We demonstrate the effectiveness of \method{} on three 7-8B scale LLMs across different model families that are adept at coding, namely, Llama3 8B Instruct~\citep{llama3modelcard}, Llama3.1 8B Instruct~\citep{dubey2024llama}, Qwen 2.5 Coder 7B Instruct in addition to the stronger Qwen 2.5 Coder 32B Instruct model~\citep{hui2024qwen2}.
\vspace{-0.5em}

\paragraph{Datasets.} We use debugging datasets based on popular LLM coding benchmarks, HumanEval~\citep{chen2021evaluating} and MBPP~\citep{austin2021program} along with their extended evaluation suites with more unit tests as proposed by \citep{liu2024your}. We describe their construction below, with further details in \cref{app:dataset}. 

\begin{itemize}[topsep=-1pt,noitemsep,leftmargin=*]
\item \textbf{HE+Fix.} We take HumanEvalFix \citep{muennighoff2024octopack}, which has human-introduced errors,
and following \citet{ninext}, augment each problem's test suite with overlapping problems from EvalPlus~\citep{liu2024your}. 
This yields 158 problems, each with a faulty solution and a suite of private UTs for evaluation -- i.e., UTs not used for debugging.  

\item \textbf{MBPP+Fix.} We construct a debugging dataset based on the MBPP+  benchmark~\citep{austin2021program,liu2024your, ninext}. 
To obtain faulty codes, we sample 16 solutions for each problem from different 7-8B scale models and filter for incorrect solutions based on the private unit tests (described in \cref{app:dataset}).
Then, we sample one faulty solution for each problem, resulting in 325 problems with realistic incorrect solutions generated by LLMs.
We find such errors are generally harder for models to debug (cf. \cref{sec:results}).

\item \textbf{MBPP+Fix (Hard).} Additionally, we create a different split from MBPP+ with more subtle errors, logical flaws and missing corner cases making it harder to debug. To this end, 
we follow a similar generation and filtering process as above but only retain faulty code that passes between $50\% \text{ - } 95\%$ of unit tests, resulting in 170 problems with faulty codes.
\end{itemize}
\vspace{-0.5em}

\paragraph{Evaluation Metrics.} Below we describe three intrinsic evaluation metrics for unit test generation (with additional details in \cref{app:eval}): 

\begin{itemize}[topsep=-1pt,noitemsep,leftmargin=*]
\item \textbf{Attack Rate.} We measure the attacking ability of a UT generator by the frequency with which the output of a gold solution $f_r$ for input $\widehat{x}$ differs from that of the buggy code $\widehat{f_b}$.

\item \textbf{Output Accuracy.} Measures how often the output of a generated unit test $\widehat{y}$ is consistent with the problem description, i.e., generates the same output as the reference code $f_r$. 

\item \textbf{Accuracy $\cap$ Attack.}  This metric combines both attack rate and output accuracy and represents how often a unit test generator $T_\theta$ generates a useful (i.e., \emph{failing}) unit test for a given target code $\widehat{f}_b$ while \emph{also} predicting the output \emph{correctly}. 
\end{itemize}

\noindent We rely on intrinsic metrics during development on a validation split from MBPP+ data, e.g., for checkpoint selection and prompt design. Then, we measure the utility and effectiveness of \method{} by computing the \textbf{pass@1 code accuracy}, i.e., the percentage of codes passing \textit{all} unit tests on code debugging and generation tasks. 
\vspace{-0.5em}

\paragraph{Baselines.} We compare \method{} against the following (prompts in \cref{app:prompts}):
\begin{itemize}[topsep=-1pt,noitemsep,leftmargin=*]
\item \textbf{\underline{No UT} feedback: } In this baseline, we use self-generated or self-critique feedback, following prior work~\citep{madaan2024self,shinn2024reflexion,chen2023teaching}.
Specifically, we prompt the model to generate an explanation about the bugs in the target code in addition to determining if the target code is correct, i.e., does not have any more bugs. 

\item \textbf{\underline{Random}ly-sampled UTs: } Consistent with \citet{chencodet}, we relax the requirement for the LLM to generate failing unit tests, and prompt the model to jointly generate valid unit tests (inputs and outputs) based on the task description, irrespective of the target code, i.e., we sample $(\widehat{x}, \widehat{y}) \sim T_\theta(d)$. We further expand on the comparison with \citet{chencodet} in \cref{ssec:ablate}. 

\item \textbf{\underline{Prompted} Failing UTs: } Using the same prompts as \method{} without any training, we prompt an LLM to generate a failing UT (with a correct output) given the description and the target code. 
Here, we sample $(\widehat{x}, \widehat{y}) \sim T_\theta(d, \widehat{f}_b)$. 
\end{itemize}

\noindent Note that all UT-generation baselines along with \method{} utilize the test-time scaling and backtracking approaches of \debugmethod{} highlighted in \cref{ssec:debug} and exhibit similar run times.

\vspace{-0.5em}
\section{Results and Analysis}
\label{sec:results}
\vspace{-0.25em}

\vspace{-0.25em}
\subsection{Intrinsic Evaluation of UT Generation}
\label{ssec:intrinsic}
\vspace{-0.25em}

In order to study the inherent UT generation abilities of different models and baselines, we use the intrinsic metrics defined by RQ1 and outlined in \cref{sec:setup} in \cref{tab:main-intrinsic} on the most challenging debugging task MBPP+Fix (Hard) averaged over 3  runs. 

\vspace{-0.25em}
\paragraph{Trade-offs in Attack Rate and Output Accuracy.} 
\begin{wraptable}{r}{0.6\linewidth}
    \vspace{-1em}
    \small
    \centering
    \setlength{\tabcolsep}{1.5pt}

    \begin{NiceTabular}{l c c c c}
    \toprule
     \bf Model  & \bf Method & \bf Attack Rate & \bf Out Acc. & \bf Acc. $\cap$ Attack \\
        \midrule
      \multirow{3}{1.25cm}{Llama3
      \\[.2\baselineskip]8B}  & Random & 30.98 & 46.86 & 10.24\\
      & Prompted & 38.04 & 37.10 & 11.51 \\
      & \method{} & {\bf 41.18} & {\bf 48.24} & {\bf 16.57} \\
        \midrule
    \multirow{3}{1.25cm}{Llama3.1\\[.2\baselineskip]8B}  & Random & 21.76 & 39.80 & 9.02\\
      & Prompted & 40.59 & 26.86 & 9.61 \\
      & \method{} & {\bf 41.37} & {\bf 47.75} & {\bf 16.67}\\
    \midrule
    \multirow{3}{1.25cm}{Qwen2.5\\[.2\baselineskip]7B}  & Random & 26.27 & 57.25 & 13.53\\
      & Prompted & 39.80 & 40.78 & 14.71\\
      & \method{} & {\bf 41.96} & {\bf 58.04} & {\bf 20.20} \\
    \midrule
    \multirow{3}{1.25cm}{Qwen2.5\\[.2\baselineskip]32B}  & Random & 25.29 & 47.25& 11.37 \\
      & Prompted & 51.96 & 48.43 & 29.41 \\
      & \method{} & \bf56.08 & \bf59.22 & \bf34.71\\
    \bottomrule
    \end{NiceTabular}
    \vspace{-0.5em}
    \caption{\textbf{Evaluation on intrinsic metrics} of different UT generation baselines and \method{} across 7-8B different model families on MBPP+Fix (Hard) over 3 runs. Higher is better for all three intrinsic metrics.}
    \label{tab:main-intrinsic}
\end{wraptable}
In \cref{tab:main-intrinsic}, we benchmark the zero-shot abilities of different models on UT generation, as well as our improved training method (corresponding to RQ2).
Here, we observe a clear tradeoff: \textit{prompted and random unit test generation each optimizing output accuracy or attack rate at the cost of the other metric}.
While randomly sampled UTs have relatively higher output accuracy (i.e., the output of the unit test is correct according to the problem description), the random baseline often lacks the ability to generate UTs with inputs that trigger errors (failing UTs). 
This can be explained by the fact that it is not conditioned on the faulty code $\widehat{f}_b$. 
On the other hand, when models are prompted to generate a UT that breaks the faulty code in the prompted UT baseline, they generally achieve a higher attack rate; however, they lag in terms of output accuracy. 
\edited{For instance, in the case of Qwen2.5 7B, switching from random sampling to prompting failing UTs improves the attack rate by 13.53\% but decreases the output accuracy by 16.47\%, hence leading to comparable Acc. $\cap$ Attack scores (13.53\% vs 14.71\%).}
Taken together, this suggests the presence of competing desirable traits of UT generators (cf. \cref{ssec:setup}; RQ1): without training, one baseline often generates trivial unit tests for which the output can easily be predicted (high output accuracy, low attack rate), while the other generates challenging unit tests for which it cannot produce the expected answer (low output accuracy, high attack rate).
It also indicates that while models can generate challenging unit test inputs that result in code failure, \emph{models struggle with reasoning over correct outputs of failing UT inputs}.

\vspace{-0.25em}
\paragraph{\method{} is Most Effective at Unit Test Generation.} 
We propose to break through this trade-off (thereby addressing RQ2, \emph{``How can we improve LLMs' UT generation abilities?''}) by training models with \method{}, where we directly supervise models to have both high output accuracy and high attack rate.
Ideally, \method{} should lead to models that generate unit tests in the ``sweet spot'' of difficulty, where they are hard enough to trigger errors but not so hard that their output cannot be predicted. 
\cref{tab:main-intrinsic} shows that models trained with \method{} consistently rank highest at jointly generating \emph{failing UT inputs with correct outputs} (as measured via Acc. $\cap$ Attack in \cref{tab:main-intrinsic}). 
\edited{For instance, on the strongest coding model in our setting, Qwen2.5 32B, \method{} obtains the highest attack rate, and output accuracy as well as improves Acc. $\cap$ Attack score by 5.3\% (absolute) over failing UTs generated by prompting and by 23.34\% over randomly generated UTs. }
Similarly, Llama3.1 improves (in terms of attack rate) over randomly-sampled UTs by 19.61\% and the output accuracy of prompted failing UTs by 20.89\%, ultimately improving the joint Acc. $\cap$ Attack score by up to 7.65\%.
We report the intrinsic metrics for HE+Fix and MBPP+Fix in \cref{app:intrinsic}.

\vspace{-0.25em}
\subsection{Generated UTs for Code Debugging} 
\label{sec:extrinsic}
\vspace{-0.25em}

\paragraph{UTs from \method{} are Best for \debugmethod{}.} Addressing RQ3, we measure how effective each type of UT generator is in a downstream debugging evaluation. From \cref{tab:extrinsic},
\begin{wraptable}{r}{0.5\linewidth}
    \small
    \centering
     \setlength{\tabcolsep}{2pt}
   
    \begin{tabular}{l c c c c }
    \toprule
        \bf Model & \bf UT Method & {\bf HE+Fix}  & \bf MBPP+Fix & \bf (Hard)  \\
    \midrule
      \multirow{4}{1.25cm}{Llama3\\[.2\baselineskip]8B} & No UT & 27.22 & 16.31 & 11.76\\
      & Random & 51.90  &  30.46 &   17.06  
      \\
       & Prompted & 51.90 &  28.92 &  22.94\\
       & \method{} & {\bf 53.80} &  {\bf 37.54} &  {\bf 28.82}   \\
    \midrule
    \multirow{4}{1.25cm}{Llama3.1\\[.2\baselineskip]8B} & No UT & 31.65 & 10.15 & 11.18\\
    & Random & 62.03 &  33.54 & 13.53\\
       & Prompted & 56.33 &  28.00 &  24.71 \\
    & \method{} & {\bf 67.09} &  {\bf 36.92} & {\bf 28.23}\\
    \midrule
    \multirow{4}{1.25cm}{Qwen2.5\\[.2\baselineskip]7B} & No UT & 52.53 & 23.08 & 16.47\\
     & Random & 79.75 &  34.77 &  17.06 \\
       & Prompted & 75.32 &  32.92 &  24.12\\
       & \method{} & {\bf 82.91} &  {\bf 37.54} &  {\bf 29.41} \\      
    \midrule
     \multirow{4}{1.25cm}{Qwen2.5\\[.2\baselineskip]32B} & No UT & 79.11 & 39.08 & 32.94 \\
     & Random & 84.81 & 49.54 &   22.94 \\
       & Prompted & 85.44 & 50.77 &  40.59\\
       & \method{} & \bf88.61 & \bf 54.15 & \bf{45.29} \\  
    \bottomrule
    \end{tabular}
    \vspace{-0.75em}
     \caption{\textbf{Evaluating pass@1 accuracies after debugging with \debugmethod{}}, using UTs generated by \method{} and other baselines for 3 rounds, on HumanEval+Fix (HE+Fix), MPBB+Fix, and MBPP+Fix (Hard). 
    }
        \label{tab:extrinsic}
\vspace{-0.75em}
\end{wraptable}
 we observe that across different LLM families, multi-turn debugging with \debugmethod{} is more effective when using generated UTs from \method{} than debugging with UTs generated by the baselines, as well as debugging without feedback. 
\edited{For instance, with Qwen2.5 32B, after 3 rounds of debugging on MBPP+Fix, \method{} improves over debugging with  randomly-sampled UTs by 4.61\%, debugging by failing UTs by 3.38\%, and debugging without UT feedback by 15.07\%. Moreover, on the more challenging MBPP+Fix (Hard) split, \method{} improves over randomly-sampled UTs by 22.35\% and the baseline without UT feedback by 12.35\%. }
Given that we use the same underlying LLM and similar feedback templates, the results in \cref{tab:extrinsic} show that UTs generated by \method{} provide the \emph{most useful and effective} feedback. 
Lastly, we find that LLM debugging without any UT feedback is least effective across model families, thus, establishing that despite noise and issues discussed in \cref{ssec:debug} and \cref{fig:fig3}, \debugmethod{} with UT-based feedback is more effective than self-generated feedback.
This aligns with past work indicating that LLMs are poor at self-critiquing~\citep{huang2024large}. 

\vspace{-0.5em}
\paragraph{Debugging Difficulty Varies Across Datasets.} 
Comparing the post-debugging accuracy across the three datasets in \cref{tab:extrinsic} suggests different difficulty levels when it comes to LLM debugging. 
The relatively high accuracies on HE+Fix (especially with Qwen2.5) suggest that human-introduced errors in that dataset are obvious and thus too easy for the models to discover and debug. 
This is also corroborated by the fact that starting codes in this dataset have a pass rate of 0\%, i.e., fail on \emph{all} unit tests, suggesting high-level coarse-grained errors, e.g., syntax errors that prevent execution, or incorrect function references. 
\edited{On the other hand, MBPP+Fix (Hard) (with an initial pass rate of $\approx$75\%) appears to be the hardest to debug, with the lowest overall post-debugging accuracy across models -- up to a difference of 53.5\% in final accuracies of HumanEvalFix and MBPP+Fix (Hard) for \method{} based on Qwen2.5 7B.}
This in turn suggests that LLMs still struggle with identifying and fixing subtle flaws in generated code, especially in scenarios involving corner cases wherein the model passes some unit tests and fails at the rest. 

\vspace{-0.5em}
\subsection{Generated UTs for Code Generation}
\label{ssec:best}
\vspace{-0.25em}

\begin{wraptable}{r}{0.48\textwidth}
    \vspace{-1em}
    \small
    \centering
    \setlength{\tabcolsep}{2.5pt}

    \begin{tabular}{l c c c}
    \toprule
     \bf Model & \bf Judge Method   & \bf HE+ & \bf MBPP+ \\
    \midrule
     \multirow{5}{*}{\rotatebox[origin=c]{90}{Llama3 8B}}    &  -$^\dagger$ & 54.43 & 56.00\\
     & RM~\citep{liu2024skywork} & 60.76 & 59.47\\
     & Random UT& 61.39 & 58.93\\
     & Prompted UT& \underline{63.29} & \underline{60.00}\\
     & \method{} & \bf 65.12 & \bf 61.60\\
     \midrule
    \multirow{5}{*}{\rotatebox[origin=c]{90}{Qwen2.5 7B}}    &  -$^\dagger$ & 81.64 & 65.06\\
     & RM~\citep{liu2024skywork} & 83.54 & 66.67\\
     & Random UT& 85.44 & \underline{68.80}\\
     & Prompted UT& \underline{85.71} & 67.47 \\
    & \method{} & \bf 87.97 & \bf 70.13\\
    \bottomrule
    \end{tabular}
    \vspace{-0.75em}
    \caption{\textbf{Best-of-10 code accuracy after judging code correctness} via generated UTs and external reward model (RM). $^\dagger$Code accuracy after sampling 1 solution.}
    \label{tab:best}
    \vspace{-1em}
\end{wraptable}
We further demonstrate the effectiveness of generated UTs at judging code correctness via Best-of-$N$  sampling ($N=10$). 
Given a problem $d$, we sample $N$ different code solutions,
 and for each solution, we generate 3 UTs (via \method{} and baselines in \cref{sec:setup} with test-time scaling). 
In the end, we collate all generated UTs from each method (removing any duplicates) and choose the generated code solution that passes the most generated UTs. 
In other words, we use \method{} to rerank 10 generated solutions. 
In addition to comparing to the random and prompt UT generation baselines, we also rerank according to
a state-of-the-art 8B scale reward model~\citep{liu2024skywork} on HE+ and MBPP+ (two standard code-generation datasets). Results in \cref{tab:best} reveal that:
\begin{itemize}[topsep=1pt,noitemsep,leftmargin=*]
\item \textbf{Generated UTs Outperform External RMs.} In all four settings (models and datasets) in \cref{tab:best}, we find that LLM-generated UT are better judges for selecting code solutions with higher pass rates as compared to trained,  state-of-the-art 8B RM.
\item \textbf{\method{} is a Better Judge for Code.} Across models and datasets, using UTs from \method{} to select code yields the highest performance, outperforming external RM by 4.36\%, randomly generated UTs as employed by \citet{chencodet} by 3.73\%, and prompted UTs by 1.83\% on HE+ with Llama3 8B (with similar trends on MBPP+ and using Qwen2.5 7B). 
\end{itemize}

\vspace{-0.5em}
\subsection{Comparison to Unit Test Generation by Frontier LLMs} 
\vspace{-0.25em}
\label{app:frontier}
\begin{wraptable}{r}{0.6\textwidth}
    \vspace{-1em}
    \small
    \centering
    \setlength{\tabcolsep}{1.5pt}

    \begin{NiceTabular}{l c c c c}
    \toprule
     \bf Model  & \bf Method & \bf Attack Rate & \bf Out Acc. & \bf Acc. $\cap$ Attack \\
     \midrule
    \multirow{2}{*}{GPT-4o} & Random & 23.92 & 57.25 & 17.25 \\
    & Prompted & 54.51 & \bf 60.39 & 33.33 \\
    \midrule
    \multirow{2}{1.25cm}{DeepSeek\\[.2\baselineskip]V3} & Random & 24.31 & 63.53 & 17.84\\
       & Prompted & 54.31 & 58.04 & 33.33\\
    \midrule
    \multirow{3}{1.25cm}{Qwen2.5\\[.2\baselineskip]32B}  & Random & 25.29 & 47.25& 11.37 \\
      & Prompted & 51.96 & 48.43 & 29.41 \\
     & \method{} & \bf56.08 & 59.22 & \bf34.71\\
    \bottomrule
    \end{NiceTabular}
     \caption{Evaluation of frontier models on intrinsic  metrics on MBPP+Fix (Hard) over 3 runs.}
    \label{tab:closed-source}
    \vspace{-1em}
\end{wraptable}
In \cref{tab:main-intrinsic} we evaluated open-source LLMs that we trained using \method{}, and in \cref{ssec:debug} we applied these models in our \debugmethod{} framework for debugging, finding the trained models to be effective. 
\cref{tab:closed-source} compares stronger frontier models -- GPT-4o~\citep{hurst2024gpt} and DeepSeek-V3~\citep{guo2024deepseek}  -- to the best open-source model tested (Qwen2.5 32B). 
Here, we see that even these stronger and much larger frontier models struggle at the task of generating failing unit tests.  
First, we find that when prompting models with faulty codes, the attack rate of the generated UTs is slightly over 50\% on MBPP+Fix (Hard) dataset, showcasing the inherently challenging nature of isolating corner cases and subtle errors in partially correct code.
\edited{Furthermore, \method{} beats both GPT-4o and DeepSeek-V3 in terms of Acc. $\cap$ Attack, with a gain of $1.38\%$ over both, as well as in terms of Attack Rate. 
On output accuracy, \method{} is better than DeepSeek-V3 and slightly worse than GPT-4o.} 
Overall, we find that using UT-based feedback even with frontier LLMs only triggers and identifies errors 1 in 3 times, leaving tremendous room for growth in the ability of models to identify partially incorrect code. 

Note that cost is a major factor in automated debugging; we show that using \method{} for debugging in \debugmethod{} is successful, but involves generating multiple UTs with self-consistency across several rounds of debugging for each problem (cf. \cref{alg:gen,alg:debug}).
Such calls will quickly become costly on frontier models like GPT-4o and DeepSeek-V3, making relatively smaller models trained with \method{} more attractive options. 

\vspace{-0.5em}
\subsection{\cameraready{Importance of Unit Tests with a Frontier Debugger}}
\label{ssec:frontier}
\vspace{-0.25em}
\begin{wraptable}{r}{0.55\textwidth}
\vspace{-0.25cm}
\small
\centering
\setlength{\tabcolsep}{1.5pt}
\begin{tabular}{lcc}
\toprule
\bf Debugger & \bf UT Method (Model) & \bf MBPP+Fix (Hard) \\
\midrule
\multirow{4}{*}{GPT4o} & No UT (GPT-4o) & 34.71 \\
 & No UT (Qwen 32B) & 32.94 \\
 & Prompted (Qwen 32B) & 45.88 \\
 & UTGen (Qwen 32B) & \textbf{59.69} \\
\bottomrule
\end{tabular}
\caption{Performance of GPT-4o as a debugger on MBPP+Fix (Hard) with different UT-generation methods. High-quality UTs from \method{} significantly improve the performance of GPT-4o.}
\label{tab:debug4o}
\vspace{-0.25cm}
\end{wraptable}
 \cameraready{Finally, we study the importance of unit tests when using a more powerful, frontier model for debugging. To this end, we use GPT-4o~\citep{hurst2024gpt} as the debugger and evaluate its performance on the MBPP+Fix (Hard) benchmark. We compare a baseline that \emph{does not} use UT-execution feedback, where GPT-4o generates its own feedback, or uses feedback generated by Qwen2.5 32B (without any unit tests) against scenarios where it uses UT feedback from the smaller Qwen2.5 32B model, both from a prompted baseline and from our proposed \method{}. The results, presented in \cref{tab:debug4o}, show that even with a significantly stronger debugger model, feedback from \textit{high-quality unit tests remains crucial} for effective debugging -- both the prompted baseline and \method{} outperform the ``no UT'' baselines by 11.17\% and $\approx$25\% (absolute), respectively. This demonstrates not only that UT feedback is vital even for frontier models, but also that the quality of the unit tests is a key factor, as \method{} outperforms the prompted baseline by a large margin of 13.8\%. }

\vspace{-0.5em}
\section{Conclusion}
\vspace{-0.25em}

We first identified a key trade-off between attack rate and output prediction accuracy when predicting unit tests with models. 
In light of this trade-off, we introduced \method{}, a new method for creating training data and teaching models to produce unit tests. 
This allows us to train models to produce better unit tests, as measured by intrinsic metrics like attack rate and output accuracy. 
Moreover, finding that existing data contains large numbers of easy errors, we introduce a new subset of data with challenging and hard-to-diagnose errors. 
To enable debugging with automated unit tests, we propose \debugmethod{}, wherein we augment our predictor's accuracy with test-time scaling and regularize it using a cross-validation and back-tracking procedure that prevents it from overfitting to a narrow or incorrect unit test.  Additionally, in \cref{app:scale}, we 
demonstrate that these gains persist as we scale the
number of generated UTs across datasets.
This, combined with \method{}, results in consistent increases in debugging performance across models, and can be used to improve code generation via best-of-N sampling. 
Finally, we note that \method{} is complementary to work on handling real-world programming issues such as \citet{swe_bench}, which focus on debugging GitHub issues raised by people; in the future, such issues could be identified by a unit test generator such as \method{}.

\vspace{-0.5em}
\section*{Acknowledgments}
\vspace{-0.25em}
\cameraready{We thank Swarnadeep Saha for insightful discussions in the early phases of this work. Furthermore, we thank the anonymous reviewers and area chairs for their helpful feedback throughout the reviewing process.
This work was supported by NSF-CAREER Award 1846185, DARPA ECOLE Program No. HR00112390060, NSF-AI Engage Institute DRL-2112635, and the Center for AI Safety Compute Cluster. 
Any opinions, findings, and conclusions or recommendations in this work are those of the author(s) and do not necessarily reflect the views of the sponsors.}

\bibliography{colm2025_conference}
\bibliographystyle{colm2025_conference}

\appendix
\section{Experimental Setup Details}
\subsection{Debugging Datasets}
\label{app:dataset}
We use debugging datasets based on popular LLM coding benchmarks, HumanEval~\citep{chen2021evaluating} and MBPP~\citep{austin2021program} along with their extended evaluation suites with more unit tests as proposed by \citep{liu2024your}.

\begin{itemize}[topsep=0pt,leftmargin=*]
\item \textbf{HE+Fix.} We start with the HumanEvalFix dataset containing human-introduced errors to gold solutions proposed by \citet{muennighoff2024octopack} (released under MIT license). 
However, this dataset only uses minimal unit tests to evaluate code correctness (from HumanEval) which has shown to be unreliable~\citep{li2023starcoder}, as it can miss errors due to low coverage. 
Therefore, we filter for problems that overlap with EvalPlus~\citep{liu2024your} released under Apache-2.0 license, which contains over 80$\times$ more unit tests. 
This yields 158 problems that each have a faulty solution and an expanded suite of private UTs -- i.e., UTs not used for debugging -- for evaluation. 

\item \textbf{MBPP+Fix.} We construct a debugging dataset based on the MBPP+  benchmark~\citep{austin2021program,liu2024your, ninext}. 
To obtain faulty codes, we sample 16 solutions for each problem from different 7-8B scale models and filter for incorrect solutions based on the private unit tests (described in \cref{app:dataset}).
Then, we select one faulty solution at random corresponding to each problem. 
This results in a total of 325 problems with incorrect solutions representing realistic errors made by LLM coders.
We find such errors are generally more challenging for models to debug (cf. \cref{sec:results}).

\item \textbf{MBPP+Fix (Hard).} To make the debugging task more challenging, we create a different split from MBPP+ with more subtle errors that are harder to debug. 
We identify these subtle errors by following a similar code generation setup described above, but only retain faulty code that passes between $50\% \text{ - } 95\%$ of unit tests, as these partially correct solutions contain less obvious logical flaws and often require handling difficult corner cases. 
This results in a total of 170 problems with faulty codes.
\end{itemize}

\textbf{HE+Fix.} This dataset contains a total of 158 problems each with one incorrect human-written code and has an initial pass rate (prior to debugging) of 0\%, i.e., all private unit tests are failing. 
As mentioned in \cref{sec:setup}, we use the dataset provided by \citet{muennighoff2024octopack}\footnote{\url{https://huggingface.co/datasets/bigcode/humanevalpack}} but replace the test set for each problem from the original test suite in HumanEval~\citep{chen2021evaluating} to that in the EvalPlus evaluation suite~\citep{liu2024your}. 
This increases the average unit tests per problem from 8.17 to 775.18 gold unit tests. 
Note that we have an automatic UT extraction script for the test code in EvalPlus, and we only retain problems for which this extraction is successful (158 out of 164).

\textbf{MBPP+Fix and MBPP+Fix (Hard).} We begin with 378 problems in the MBPP+ dataset~\citep{liu2024your}\footnote{\url{https://huggingface.co/datasets/evalplus/mbppplus}} and follow the same gold UT extraction step described above, discarding problems for which the extraction fails. 
This leaves us with 375 problems, for which we sample 16 solutions per problem across multiple LLMs: Gemma-7B-IT~\citep{team2024gemma}, Llama3 8B Instruct~\citep{llama3modelcard}, Llama3.1 8B Instruct~\citep{dubey2024llama}, Tulu-3 8B SFT~\citep{lambert2024tulu3}, DeepSeek 7B coder~\citep{guo2024deepseek}, Qwen2.5 Coder 7B~\citep{hui2024qwen2}. 
To generate MBPP+Fix, we filter for incorrect solutions (i.e., with at least one failing UT) and then randomly sample one incorrect code per problem. 
This yields 325 problems in total, each with one faulty code.
This dataset has an initial pass rate of 24.21\% and an average of 107.45 gold unit tests per problem. 
In order to construct, MBPP+Fix (Hard), we follow a similar process but select only incorrect solutions which pass 50 - 95\% of unit tests. 
The intuition here is that solutions that are \emph{partially} correct are often harder to debug than those that are fully incorrect. 
We then randomly sample one such incorrect solution per problem, yielding a dataset of 170 problems with an initial pass rate of 74.83\% and an average of 107.49 gold unit tests.

\paragraph{Code Generation.} When using generated UTs for improving code generation, we evaluate on HumanEval+ (HE+) and MBPP+ benchmarks proposed in \citet{liu2024your} released under Apache-2.0 license. As described above, we filter out problems for which we can extract private UTs, yielding 158 problems in HE+ dataset and 375 problems in MBPP+ dataset. For the RM baseline we use the latest version of the 8B model based on Llama-3.1 backbone in \citet{liu2024skywork}.\footnote{\url{https://huggingface.co/Skywork/Skywork-Reward-Llama-3.1-8B-v0.2}} We made this choice based on  on rankings on the RewardBench leaderboard~\citet{RewardBench}

\subsection{Evaluation Metrics}
\label{app:eval}
Below we describe three intrinsic evaluation metrics for unit test generation: \emph{attack rate, output accuracy, } and \emph{accuracy $\cap$ attack}; along with pass@1 accuracy as the extrinsic metric to measure LLM's debugging abilities using UT-feedback. 

\begin{itemize}[topsep=0pt,leftmargin=*]
\item \textbf{Attack Rate.} This metric measures a UT generator $T_{\theta}$'s attacking ability, i.e., its ability to generate a failing unit test input $\widehat{x}$ for a given buggy solution $\widehat{f_b}$. 
We measure this by matching if the output of a gold reference solution $f_r$ for input $\widehat{x}$ differs from that of the buggy code $\widehat{f_b}$. 
Note that this does not take into account the accuracy of the unit test output which we measure separately below.
Mathematically, for any dataset $\mathcal{D}$ of coding problems, attack rate is defined as: 
\vspace{-0.2cm}
\begin{align*}
\mathrm{Attack Rate} &= \frac{100}{|\mathcal{D}|} \times \sum_{d \in \mathcal{D}} \mathbbm{1}_{f_r(\widehat{x}) \neq \widehat{f_b}(\widehat{x})}; \\
&\text{where} (\widehat{x}, \widehat{y}) \sim T_\theta(d, \widehat{f_b})  
\end{align*}
\vspace{-0.2cm}

\item \textbf{Output Accuracy.} This metric measures how often the output of a generated unit test $\widehat{y}$ is consistent with the problem description, i.e., generates the same output as the reference gold code $f_r$. Output accuracy does not require the generated unit test to fail. 
In other words, 
\vspace{-0.2cm}
\begin{align*}
\mathrm{Output Acc} &= \frac{100}{|\mathcal{D}|} \times \sum_{d \in \mathcal{D}} \mathbbm{1}_{\widehat{y} = f_r(\widehat{x})}; \\ 
&\text{where} (\widehat{x}, \widehat{y}) \sim T_\theta(d, \widehat{f_b})
\end{align*}
\vspace{-0.2cm}

\item \textbf{Accuracy $\cap$ Attack.}  This metric combines both attack rate and output accuracy and represents how often a unit test generator $T_\theta$ generates a useful (i.e., \emph{failing}) unit test for a given target code $\widehat{f}_b$ while \emph{also} predicting the output \emph{correctly}. We calculate this as, 
\begin{align*}
    \mathrm{Acc. \cap Attack} = &
\frac{100}{|\mathcal{D}|}\!\!\times\!\!\sum_{d \in \mathcal{D}}\!\mathbbm{1}_{f_r(\widehat{x}) \neq (\widehat{f}_b)(\widehat{x}), \widehat{y} = f_r(\widehat{x})}\\
& \text{where } (\widehat{x}, \widehat{y}) \sim T_\theta(d, \widehat{f_b})
\end{align*}

\item \textbf{Code Accuracy.} To evaluate the utility of generated unit tests via code debugging, we follow prior work~\citep{chen2023teaching, chae2024coffee} in reporting pass@1 code accuracy, i.e., the percentage of codes passing \textit{all} unit tests after 3 rounds of debugging. Note that while we \emph{debug} with model-generated UTs, we evaluate code accuracy using private \emph{human-annotated} UTs.

\end{itemize}

\input{utdebug_alg}

\section{\method{} Training}
\label{app:method}
\textbf{Preprocessing Tulu Data.} We use the Tulu-3 SFT mixture dataset released by \citet{lambert2024tulu3} which contains a total of 939.3K instances.\footnote{\url{https://huggingface.co/datasets/allenai/tulu-3-sft-mixture}}  
However, it contains a mixture of prompts for instruction-following, math reasoning, and coding.
Therefore, we filter for prompts involving Python coding by regex search for keywords ``\texttt{python}'' and ``\texttt{def }'' which suit our task setup described in \cref{ssec:setup}. 
Furthermore, we filter out instances with more than 2K tokens in the prompt and ensure the prompt is a valid unicode string. 
This results in a total of 48.3K instances for which we use the ``prompt'' as the problem description and extract code from the response of the last turn when multi-turn interactions are present provided that the extracted code executes without any errors or issues. 
Finally, we prompt the LLM to be trained with \method{} to generate 2 corrupted versions of this code to serve as the target code, and for each target code, we make 5 attempts to generate failing unit test inputs and filter out instances that do not have any such UTs. 
This is followed by the rationalization step (with the same LLM to be trained) described in \cref{ssec:training} and in \cref{fig:method}, which results in roughly 30K instances for each 7-8B model trained with \method{} and nearly 70K instances for Qwen2.5 32B. 

\begin{table*}[!t]
    \small
    \centering
    \setlength{\tabcolsep}{1.5pt}

    \begin{tabular}{l c c c c  c c c }
    \toprule
        \bf Model & \bf UT Method & \multicolumn{3}{c}{\bf HE+Fix}  & \multicolumn{3}{c}{\bf MBPP+Fix}\\
        \cmidrule(lr){3-5}\cmidrule(lr){6-8}
        & & \bf Attack Rate & \bf Out Acc. & \bf Acc. $\cap$ Attack & \bf Attack Rate & \bf Out Acc. & \bf Acc. $\cap$ Attack \\
    \midrule
      \multirow{3}{*}{Llama3 8B} & Random & 89.63 & {\bf 72.97} & {\bf 72.97} & 62.56 & 41.85 & 24.28\\
       & Prompted & 95.73 &39.59 & 38.78 & 62.67 & 29.64 & 16.41\\
       & \method{} & {\bf 96.34} & 53.27 & 52.04 & {\bf 67.18} & {\bf 42.67} & {\bf 26.87} \\
    \midrule
    \multirow{3}{*}{Llama3.1 8B} & Random & 76.02 & {\bf 63.52} & {\bf 63.52}  & 47.28 & 36.0 & 21.33\\
       & Prompted & 92.68 & 34.53 & 33.89 & 59.28 & 19.38 & 11.08\\
       & \method{} & {\bf 96.54} & 56.38 & 55.76 & {\bf 62.77} & {\bf 43.59} & {\bf 25.54} \\
    \midrule
    \multirow{3}{*}{Qwen2.5 7B} & Random & 90.85 & {\bf 87.36} & {\bf 86.47} & 55.28 & {\bf 52.31} & 30.38  \\
       & Prompted & 93.29 & 54.55 & 53.91 & 62.97 & 35.29 & 22.60 \\
       & \method{} & {\bf 96.54} & 72.90 & 72.48 & {\bf 65.54} & 48.62 & {\bf 32.82} \\
    \midrule
    \multirow{3}{*}{Qwen2.5 32B} & Random & 84.96 & \bf 80.19 & \bf 79.95 & 54.56 &	50.17 &	34.87  \\
       & Prompted &  88.82 & 	64.77 &	63.89 & 66.87	& 35.59 &	27.38 \\
       & \method{} &  \bf 95.93 &	70.64 & 	69.79  & \bf 70.36	& \bf 50.87	& \bf 40.62\\
    \bottomrule
    \end{tabular}
    \caption{Evaluation on intrinsic metrics on HE+Fix and MBPP+Fix for different UT generation methods across multiple models.}
    \label{tab:extra-intrinsic}
\end{table*}

\paragraph{Training Hyperparameters.} We train each model for 10 epochs with a batch size of 16 and a learning rate of 5e-6 with a cosine learning rate scheduler. 
Moreover, we only compute negative log-likelihood loss over the completions. 
We use LoRA~\citep{hu2021lora} with a rank of 16, $\alpha$ = 32 with a dropout of 0.05. We perform checkpoint selection based on the intrinsic Acc. $\cap$ Attack metric. All training and inference is conducted on Nvidia's A6000 GPUs taking $\approx$20 GPU hours for training and $<0.5$ GPU hours for inference (generating one code solution and corresponding UTs with test-time scaling). We train the larger Qwen2.5 32B model on 8 A100 GPUs for roughly 40 GPU hours using QLoRa~\citep{dettmers2023qlora} with 4-bit quantization.

\section{Overall Pipeline for \debugmethod{}}
\label{app:debug}

In \cref{alg:gen} we describe the process of generating UTs for a candidate buggy code $\widehat{f}_b$ using any UT generation method and perform test time scaling for UT output prediction. 
This is then used within \debugmethod{} as shown in \cref{alg:debug}, which identifies failing UTs, debugs the target code $\widehat{f}_b$ based on feedback from failing UTs over multiple rounds, and returns the debugged (edited code). 
As illustrated in \cref{fig:fig3} (b), edits are accepted only if the newly generated code achieves a higher pass rate than the code before debugging, otherwise the edits are discarded.

\textbf{Inference Hyperparameters.} When sampling multiple UTs and for generating LLMs response to UT feedback we use temperature sampling with $\mathrm{temp}\!=\!0.7$ and $\mathrm{top\_p}\!=\!0.9$.
\section{Additional Intrinsic Evaluation}
\label{app:intrinsic}

Similar to \cref{tab:main-intrinsic}, we report intrinsic evaluation metrics (cf. \cref{sec:setup}) for HE+Fix and MBPP+Fix datasets in \cref{tab:extra-intrinsic}. 
Consistent with the findings in \cref{ssec:intrinsic}, without training, we observe a trade-off between attack rate and output accuracy, with randomly sampled UTs showing higher output accuracy whereas the prompted baseline exhibits higher attack rates and vice-versa. 
Once again \method{} training best balances this trade-off yielding both high output accuracy and attack rate. 
However, due to the relative ease of attacking faulty codes in HE+Fix (initial pass rate of 0\%) almost any generated UT fails and thus can be used for debugging. 
This, coupled with the higher output accuracy, results in the random baseline having the highest score on Acc. $\cap$ Attack. 
Note that we mitigate this impact on downstream performance by using test-time scaling for output prediction in \debugmethod{}, which especially boosts the accuracies of targeted UTs generated based on the buggy code. 
On MBPP+Fix, \method{} consistently yields the highest Acc. $\cap$ Attack score, followed by the random baseline. 

\begin{wraptable}{r}{0.4\textwidth}
   \vspace{-1em}
    \small
    \centering
    \setlength{\tabcolsep}{2pt} 
    \begin{tabular}{l c}
    \toprule
      \bf{Intrinsic Metric}   &  \bf{Somers' Delta}\\
    \midrule
      Attack Rate   & 0.43\\
      Output Acc. & 0.34\\
      Acc. $\cap$ Attack (Both) & 0.63 \\
    \bottomrule
    \end{tabular}
\caption{Somers' D correlation between intrinsic UT metrics and whether it resulted in successful debugging. Note, all reported results are statistically significant ($p < 0.05$).}
\label{tab:correl}
\end{wraptable}
\cameraready{To further validate our intrinsic evaluation metrics, specifically, that \emph{both} high input attack rate and high output accuracy are key to downstream utility of model-generated unit tests -- we measure the correlation between each property of the UT and whether the UT yielded successful debugging. To this end, we collect oracle metadata on all unit tests generated for MBPP+Fix with Qwen2.5 7B, i.e., we record if the unit test caused the target code to fail, whether its output was correct, and compute its Somers' D  correlation with whether the target code was debugged successfully for $n=1$ UTs and 1 round of debugging in \cref{tab:correl}. While individually, failing UTs and output correctness show mild correlation with downstream debugging, \cref{tab:correl} shows that the presence of both attributes yields a strong (and the highest) correlation with debugging success and reinforces the motivation for \method{} and improving unit-test generation ability of LLMs at large. }

\section{\cameraready{Additional Analysis and Results with \debugmethod{}}}
\label{ssec:ablate}

\paragraph{On Test-time Scaling and Backtracking.}
In \cref{ssec:debug}, we highlighted the challenges of debugging with imperfect UTs and suggested remedies to make our debugging pipeline, \debugmethod{}, robust to noisy feedback from such UTs. We study the effectiveness of these measures, i.e., test-time scaling of UT outputs, and backtracking,  on debugging with $n=3$ generated UTs for 3 rounds using Qwen2.5 7B on the MBPP+Fix dataset in \cref{tab:ablate}.
\begin{wraptable}{r}{0.5\textwidth}
    \vspace{-1em}
    \small
    \centering
    
    \setlength{\tabcolsep}{2pt}
    \begin{tabular}{l c c}
    \toprule
      \bf UT Method   &  \bf Acc. & $\Delta$\\
    \midrule
    Randomly-sampled (Qwen2.5 7B)& 34.77 & \\
    \qquad - Output Test-time Scaling & 30.77& \textcolor{purple}{- 4.0}\\
    \qquad - Backtracking & 32.61 & \textcolor{purple}{- 2.2}\\
    \midrule
    \method{} (Qwen2.5 7B) & 37.54 & \\
    \qquad - Output Test-time Scaling & 26.15 & \textcolor{purple}{- 11.4} \\
    \qquad - Backtracking & 34.38 & \textcolor{purple}{- 3.2}\\
    \bottomrule
    \end{tabular}
    \caption{Ablating components of \debugmethod{}'s  pipeline (cf. \cref{ssec:debug}) for two different unit test generation methods (including \method{}) on MBPP+Fix using Qwen2.5.}
    \label{tab:ablate}
    \vspace{-0.5em}
\end{wraptable}
We show that {test-time scaling and backtracking are \emph{crucial} for LLM debugging with \emph{generated} UTs.} From \cref{tab:ablate}, we observe that irrespective of the underlying method for UT generation (either randomly sampled or from \method{}), removing either backtracking or test-time scaling hurts downstream performance.
First, we find that removing test-time scaling for predicting the output of UTs decreases the performance of randomly-sampled UTs by 4\% and that of \method{} by 11.4\%. Note that a larger drop in \method{}'s performance when removing test-time scaling of UT output prediction is consistent with the findings in \cref{ssec:intrinsic} that models struggle with reasoning over correct outputs for \emph{failing UT inputs} (more often generated by \method{}). Therefore, test-time scaling in output prediction provides a greater boost for reasoning over challenging inputs~\citep{wang2022self}, and consequently, removing it yields larger drops for \method{}. Moreover, \cref{tab:ablate} demonstrates that, without validation on other generated unit tests, LLMs tend to \emph{overfit} on the unit test contained in the feedback, resulting in up to 3.2\% drop in performance of \method{} without backtracking. 

\paragraph{\cameraready{Comparison with CodeT~\citep{chencodet} for Debugging.}}
\begin{wraptable}{r}{0.575\textwidth}
\vspace{-1em}
\small
\centering
\setlength{\tabcolsep}{2pt}
\begin{tabular}{llccc}
\toprule
\textbf{Model} & \textbf{UT Method} &\textbf{ HE+Fix} & \textbf{MBPP+Fix} & \textbf{(Hard)} \\
\midrule
\multirow{5}{*}{Qwen 2.5 7B} & No UT & 52.53 & 23.08 & 16.47 \\
 & Random & 79.75 & \underline{34.77} & 17.06 \\
 & Prompted & 75.32 & 32.92 & \underline{24.12} \\
 & CodeT & \underline{81.65} & 34.30 & 21.18 \\
 & \method{} & \textbf{82.91} & \textbf{37.54} & \textbf{29.41} \\
\midrule
\multirow{5}{*}{Qwen 2.5 32B} & No UT & 79.11 & 39.08 & 32.94 \\
 & Random & 84.81 & 49.54 & 22.94 \\
 & Prompted & 85.44 & \underline{50.77} & \underline{40.59} \\
 & CodeT & \underline{86.71} & 46.61 & 29.41 \\
 & \method & \textbf{88.61} & \textbf{54.15} & \textbf{45.29} \\
\bottomrule
\end{tabular}
\caption{Comparison of pass@1 accuracies after $n=3$ rounds of debugging with \debugmethod{} of different UT-generation methods including CodeT~\citep{chencodet} -- a variant of random UT generation baseline -- as well as \method{}.}
\label{tab:codet}
\vspace{-0.5em}
\end{wraptable}
\cameraready{As mentioned in \cref{sec:rel}, \method{} differs from \cite{chencodet} in that CodeT is designed for code generation by selecting a code from the largest consensus set of codes and independently generated UT (without conditioning on code). Alternatively, this UT sampling procedure is akin to the randomly-sampled baseline (described in \cref{sec:setup}) but the latter uses self-consistency for predicting the UT output. In contrast, \method{} is designed for UT generation focusing on the \textit{intrinsic quality of UTs} as well as their \textit{utility for downstream tasks} like code-debugging and best-of-N ranking by conditioning UT generation on the edge cases of a target code. To directly compare against CodeT, instead of using the code generated from CodeT, we use its unit tests in downstream debugging.
In \cref{tab:codet}, we scale up CodeT such that it uses a similar computational budget as \method{} and sample $n=3$ unit tests for debugging with Qwen 2.5 7B and 32B Code-Instruct models. Across both models and all three debugging datasets, \method{} outperforms debugging with unit tests generated by CodeT by as much as 7.54\% (absolute) on MBPP+Fix and 15.88\% (absolute) on MBPP+Fix (Hard) with Qwen 2.5 32B model. Moreover, CodeT (which generates UTs independent of code being debugged) lags behind the prompted UT baseline in nearly half of the settings, showing that conditioning UT generation on the erroneous code better helps identify and localize bugs. 
}

\section{Scaling with Number of Generated UTs}
\label{app:scale}
\begin{figure}[ht]
    \centering
    \includegraphics[width=0.95\linewidth]{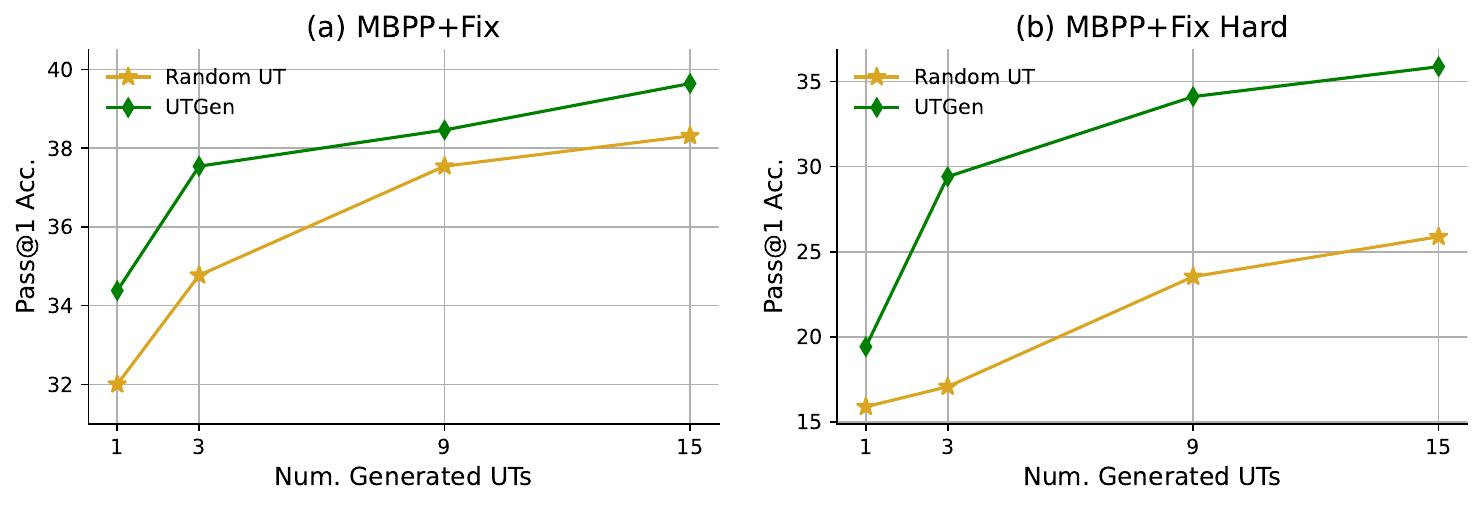}
    \caption{Increasing number of UTs across MBPP+Fix and MBPP+Fix (Hard) using UTs generated by \method{} and randomly-sampling with Qwen2.5 7B. 
    }
    \label{fig:scaling}
\end{figure}
Thus far, we use $n\!=\!3$ generated UTs across baselines and models. However, as we described in \cref{ssec:debug}, having multiple UTs can be advantageous because: (i) there is a higher likelihood of generating a failing UT and getting a reliable signal for when the code is correct, (ii) more robust signal for when to backtrack using validation on the entire generate test suite. We analyze the impact of increasing the number of generated UTs $n$ on downstream accuracy of Qwen2.5 7B after 3 rounds of debugging in \cref{fig:scaling}. 
Our findings are as follows:
\begin{itemize}[noitemsep,leftmargin=*]
\item First, \cref{fig:scaling} shows that despite increasing the number of generated unit tests, \method{} consistently outperforms randomly sampled UTs that may not be failing. This highlights the benefits of generating \emph{targeted unit tests conditioned on buggy code} in order to trigger errors and generate appropriate feedback for debugging.  

\item In settings with constrained resources, i.e., sampling $n\leq3$ UTs, \method{} is more effective at identifying errors by up to 3\% on MBPP+Fix and 12\% on MBPP+Fix (Hard). 

\item On MBPP+Fix (Hard), which contains less obvious errors and is harder to debug, we find that despite scaling to up to $15$ generated UTs, the performance gap between \method{} and randomly-sampled UTs remains at 10\% (absolute). 

\end{itemize}

\section{Discussion}
Our work on \method{} contributes to the broader landscape of verification and feedback generation for LLM-generated code. While recent work has focused on training verifiers to provide feedback~\citep{mahan2024generative,zhang2024generative}, a key challenge remains in obtaining high-quality feedback signals for debugging. 
\method{} addresses this by directly generating unit tests that can identify problems in code, complementing existing work on how to effectively incorporate and present feedback for debugging~\citep{chen2023teaching, zhong2024ldb} along with test-time scaling and backtracking incorporated in \debugmethod{}. Our results demonstrate that without a quality signal to determine code correctness and/or how a faulty code is failing (in the form of unit tests), using LLMs to generate feedback and debug still proves to be challenging. This is one of the first efforts in this direction, and we hope to spark more interest in future work toward LLM-generated unit tests (both input and outputs) that reveal the model's coding errors. 

Our approach connects to and complements recent work on handling real-world programming issues. While approaches designed for SWEBench \citep{swe_bench} focus on fixing known issues from GitHub by understanding and implementing fixes for bug reports, \method{} addresses the upstream challenge of automatically discovering potential issues in new code through test generation. 
Both tasks share a core challenge: determining the expected behavior of code without access to correct implementations. 
This connects to the fundamental concept of simulatability from computability theory \citep{rices_theorem}, where we ask whether a program can predict the behavior of another program. 
Recent work such as \citet{jain2024livecodebench} shows that while LLMs can often simulate existing code by tracing execution steps, they struggle more with predicting correct outputs from specifications alone. 
Our results align with these findings -- while \method{} can generate test inputs that trigger errors (high attack rate), predicting correct expected outputs remains challenging (lower output accuracy). 
This suggests that improving LLMs' ability to reason about intended program behavior from specifications remains a crucial direction for future work. 
Nevertheless, we find that the modifications made to debugging in \debugmethod{} help boost \method{}'s accuracy and account for noise, leading to downstream gains.

\section{Prompts}
\label{app:prompts}
In the following pages, we list all the prompts we use in this work.
\begin{user_example}[frametitle={Prompted and UTGen Prompt for UT generation}]
You are given a Python function \{signature\} to solve the following task:

\#\# Task:

\{description\}

\#\# Code Solution:

\{code\}

The code solution I have provided to you is **incorrect**. Your job is to give feedback by generating a unit test that 

1. Is **valid** input based on the task description, i.e., an acceptable input consistent with task description that a correct program should be able to execute.

2. The output enclosed in . and is **faithful** to the task description, i.e., the output of the unit test is consistent with what a correct program would return.

3. **Breaks** the given code, i.e., does **not** execute to the **correct** output and brings out its mistakes and vulnerabilities.

Provide a reasoning for your answer and identify a general hypothesis or rationale identifying the cause of error. Then provide input and output of the unit test consistent with the pattern (hypotheis) you have identified.
Note:
- that you MUST directly write ALL input arguments of the function in the correct order. Skip writing any names of arguments.
- you MUST enclose the unit test inputs and outputs in.

Respond in the format below:

\#\# Hypothesis

\textless  step-by-step reasoning  \textgreater

Error Pattern: \textless an identified pattern of inputs that yields erroneous or incorrect outputs \textgreater

\#\# Unit Test

\#\#\# Input Arguments

\textless  step-by-step reasoning for constructing a unit test that fits the error pattern identified above and is valid as per the task description \textgreater
Arguments: \{entry\_point\}(\textless all arguments \textgreater)

\#\#\# Output

\textless step-by-step reasoning for what a **correct** \{entry\_point\} would execute to based on the task description and your input above. Make sure your data type of the final answer matches the expected output type of the function.  \textgreater

Output: \textless your final answer \textgreater

\end{user_example}
 \clearpage
\begin{user_example}[frametitle={No UT Feedback Prompt}]
\# Your Task

\#\# Task
\{prompt\}

\#\# Code:

\{code\}

Based on given task and code, generate feedback that decides whether the code is correct or wrong in the format Feedback: \textless your feedback \textgreater.
Always end your feedback with the line ``The above code is correct.'' if it is correct, otherwise the feedback should end with ``The above code is wrong, please fix it.''
\end{user_example}

\begin{user_example}[frametitle={UT Generation Prompt for Randomly-sampled UTs}]
Given a Python function \{signature\} to solve the following task:
\{description\}

Code Solution:

\{code\}

The code solution I have provided to you is incorrect. Your job is to give feedback by generating a unit test input that 
1. Valid input based on the task description, i.e., an acceptable input consistent with task description that a correct program should be able to execute.
2. Given code will NOT be able to solve and brings out its mistakes and vulnerabilities.

Provide a reasoning for your answer and present your response in the format below:

\textless reasoning \textgreater

Arguments: \{entry\_point\}(\textless all arguments \textgreater)

Note that you MUST directly write ALL input arguments of the function in the correct order. Skip writing any names of arguments.

\end{user_example}

\begin{user_example}[frametitle={Training UT Gen -- Code Corruption Prompt}]
You are given a Python function \{signature\} to solve the following task:

\#\# Task

\{description\}

\#\# Correct Code Solution:

\{code\}

Assume you are a TA for a programming course. Your task is to corrupt the correct code or implementation of function \{entry\_point\} to introduce realistic errors that can be made by your programming students.
Note that you should write a code that fails one or more unit tests that this correct would succeed in.
Also, give all your reasoning for why your generated code is incorrect outside the code block, i.e., **do not** leave comments in the code block that reveals the code is incorrect. 

Give your output in the format:

\textless reasoning of error introduced \textgreater

\#\# Incorrect Code Solution

\textless your generated incorrect code for \{entry\_point\} \textgreater

\end{user_example}

\begin{user_example}[frametitle={UT Feedback Prompt}]
The above code is incorrect and does not pass the testcase.

Input: \{wrong\_testcase\_input\}

Output: \{wrong\_testcase\_output\}

Expected: \{wrong\_testcase\_expected\}
\end{user_example}

\begin{user_example}[frametitle={Output Rationalization Prompt }]
\#\# Example

Given a Python function check(string) to solve the following task:

Write a python function to accept the strings which contains all vowels.

A user provides an the following input to function. The teacher lets you know that correct function generates the following output.

Input: "BCDEFG"

Output: False

Now **without** coding up the the function check(string), provide step-by-step reasoning for why function check when given input of "BCDFG" generates or returns False. 

\#\#\# Reasoning

Let's think step by step.

- According to the problem description, given a string the check should return accepted if it contains all the vowels and not accepted otherwise.

- The vowels (in lower case) that should be present in the string are: "a", "e", "i", "o" and "u".

- The given input is "BCDEFG" which contains characters: "b", "c", "d", "e", "f", "g".

- While the input string "BCDEFG" contains only vowel "e" and is missing vowels: "a", "i", "o", "u".

- Therefore, the output of the function is not accepted.

\#\# Test Problem

Given a Python function \{signature\} to solve the following task:

\{description\}

A user provides an the following input to function  \{signature\}. The teacher lets you know that correct function generates the following output.

Input: \{unit\_input\}

Output: \{unit\_output\}

Now **without** coding up the the function \{signature\}, provide step-by-step reasoning for why function \{entry\_point\} when given input of \{unit\_input\} generates or returns \{unit\_output\}.

Write your output under header \#\#\# Reasoning.
\end{user_example}

\end{document}

%% file: utdebug_alg.tex
\begin{algorithm}[t]
\caption{$\tt{BuildUT}$: Build Generated Unit Test Suite}
\label{alg:gen}
\begin{algorithmic}{}
    \State {\bfseries Input:} $d, \hat{f}_b$ \textcolor{comm}{\footnotesize{\textit{// problem description and buggy code}}}
    \State {\bfseries Params:}  Number of UTs $n$, Number of SC samples $k$, Unit Test Generator $T_\theta$
    \State {\bfseries Output:} Set of Generated UTs $\mathcal{U}$
    \State $\mathcal{U} \gets \varnothing$  \textcolor{comm}{\footnotesize{\textit{// Initialization, i.e., no generated UTs}}} 
    \For{ UT index $i$ $\in [1, \cdots, 3\times n]$}
        \State \textcolor{comm}{\footnotesize{\textit{// Generates up to $n$ distinct UTs from the UT generator $T_\theta$}}} 
        \State $(\hat{x}^i, \hat{y}^i) \sim T_\theta(d, \hat{f}_b)$ \textcolor{comm}{\footnotesize{\textit{// Sample UTs from UT generator}}}
        \State $\nu^i \gets \varnothing$  \textcolor{comm}{\footnotesize{\textit{// Initialize vote lookup for self-consistency}}} 
        \For{Output index $j \in [1, \cdots, k]$}
        \State $r^i_j \sim T_\theta(d, \hat{f}_b | x^i)$ \textcolor{comm}{\footnotesize{\textit{// Sample UT output}}}
        \State $y_j \gets \tt{extractAns}(r^i_j)$ \textcolor{comm}{\footnotesize{\textit{// Extract UT output}}}
        \State $\nu^i[y_j] \gets \nu^i[y_j] + 1$ \textcolor{comm}{\footnotesize{\textit{// Append vote tally}}}
        \EndFor
        \State $\hat{y}^i \gets \mathrm{arg max}(\nu^i)$
        \textcolor{comm}{\footnotesize{\textit{// Use majority vote for UT output}}}
        \If{$\nu^i[\hat{y}^i] \geq 0.5k$}
        \textcolor{comm}{\footnotesize{\textit{// Answer gets over 50\% of the vote}}}
        \State $\mathcal{U} \gets \mathcal{U} + (\hat{x}^i, \hat{y}^i)$ \textcolor{comm}{\footnotesize{\textit{// Add to generated UT set}}}
        \EndIf
        \If{ $|\mathcal{U}| \geq n$}  \Return{$\mathcal{U}$}
        \EndIf
    \EndFor
    \Return{$\mathcal{U}$}
\end{algorithmic}

\end{algorithm}

\begin{algorithm}[!t]
\caption{$\tt{UTDebug}$: Debugging with generated UTs}
\label{alg:debug}
\begin{algorithmic}{}
    \State {\bfseries Input:} $d, \hat{f}_b$ \textcolor{comm}{\footnotesize{\textit{// problem description and buggy code}}}
    \State {\bfseries Params:}  Number of UTs $n$, Number of SC samples $k$, Unit Test Generator $T_\theta$, Number of debugging rounds $m$
    \State {\bfseries Output:} Debugged code $\hat{f}_e$
    \State $i \gets m$  \textcolor{comm}{\footnotesize{\textit{// Initializing number of rounds left}}}
    \State $\hat{f}_e \gets \hat{f}_b$  \textcolor{comm}{\footnotesize{\textit{// Initializing edited code with code to debug}}} 
    \State $\mathrm{acceptEdit} \gets True$  \textcolor{comm}{\footnotesize{\textit{// Accept edits to start debugging}}} 
    \While{$i > 0$}
    \State $i \gets i - 1$ \textcolor{comm}{\footnotesize{\textit{// One round of debugging}}} 
    \If{$\mathrm{acceptEdit} = True $}
    \State $\mathcal{U} \gets {\tt{BuildUT}}(d, \hat{f}_e)$ \textcolor{comm}{\footnotesize{\textit{// Obtain generated UTs}}}
    \State $(x_d, y_d) \gets \varnothing$ \textcolor{comm}{\footnotesize{\textit{// Initialize UT for debugging feedback}}}
    \For{$(x,y) \in \mathcal{U}$}
    \If{$\hat{f}_e(x) \neq y$}
    \State $(x_d, y_d) \gets (x,y) $  \textcolor{comm}{\footnotesize{\textit{// Failing UT  to debug}}}
    \EndIf
    \State $\mathrm{prePass} \gets {\tt{EvalCode}}(\hat{f}_e, \mathcal{U})$ \textcolor{comm}{\footnotesize{\textit{ // Get pass rate}}} 
    \If{$x_d = \varnothing$} \textcolor{comm}{\footnotesize{\textit{// No need to debug}}} 
    \State \Return{$\hat{f}_e$}
    \Else \textcolor{comm}{\footnotesize{\textit{ // Prompt LLM to debug code}}} 
    \State $f'\sim{\tt{LLM}}(\hat{f}_e | {\tt Debug}(x_d, y_d, \hat{f_e}))$
    \textcolor{comm}{\footnotesize{\textit{ // Prompt LLM to debug code with UT feedback}}} 
    \State $\mathrm{postPass} \gets {\tt{EvalCode}}(f', \mathcal{U})$ 
    \If{$\mathrm{postPass} > \mathrm{prePass}$}
    \State $\hat{f}_e \gets f'$  \textcolor{comm}{\footnotesize{\textit{ // Based on validation on the generated UTs, accept the edits, otherwise discard, i.e., backtrack}}} 
    \EndIf
    \EndIf
    \EndFor
    \EndIf
    \EndWhile
    \Return{$\hat{f}_e$}
\end{algorithmic}
\end{algorithm}

%% file: colm2025_conference.bbl
\begin{thebibliography}{56}
\providecommand{\natexlab}[1]{#1}
\providecommand{\url}[1]{\texttt{#1}}
\expandafter\ifx\csname urlstyle\endcsname\relax
  \providecommand{\doi}[1]{doi: #1}\else
  \providecommand{\doi}{doi: \begingroup \urlstyle{rm}\Url}\fi

\bibitem[Achiam et~al.(2023)Achiam, Adler, Agarwal, Ahmad, Akkaya, Aleman, Almeida, Altenschmidt, Altman, Anadkat, et~al.]{achiam2023gpt}
Josh Achiam, Steven Adler, Sandhini Agarwal, Lama Ahmad, Ilge Akkaya, Florencia~Leoni Aleman, Diogo Almeida, Janko Altenschmidt, Sam Altman, Shyamal Anadkat, et~al.
\newblock Gpt-4 technical report.
\newblock \emph{arXiv preprint arXiv:2303.08774}, 2023.

\bibitem[AI@Meta(2024)]{llama3modelcard}
AI@Meta.
\newblock Llama 3 model card.
\newblock 2024.
\newblock URL \url{https://github.com/meta-llama/llama3/blob/main/MODEL_CARD.md}.

\bibitem[Anthropic(2024)]{claude3modelcard}
Anthropic.
\newblock The claude 3 model family: Opus, sonnet, haiku.
\newblock 2024.
\newblock URL \url{https://www-cdn.anthropic.com/de8ba9b01c9ab7cbabf5c33b80b7bbc618857627/Model_Card_Claude_3.pdf}.

\bibitem[Austin et~al.(2021)Austin, Odena, Nye, Bosma, Michalewski, Dohan, Jiang, Cai, Terry, Le, et~al.]{austin2021program}
Jacob Austin, Augustus Odena, Maxwell Nye, Maarten Bosma, Henryk Michalewski, David Dohan, Ellen Jiang, Carrie Cai, Michael Terry, Quoc Le, et~al.
\newblock Program synthesis with large language models.
\newblock \emph{arXiv preprint arXiv:2108.07732}, 2021.

\bibitem[Beck(2022)]{beck2022test}
Kent Beck.
\newblock \emph{Test driven development: By example}.
\newblock Addison-Wesley Professional, 2022.

\bibitem[Cadar et~al.(2008)Cadar, Dunbar, Engler, et~al.]{cadar2008klee}
Cristian Cadar, Daniel Dunbar, Dawson~R Engler, et~al.
\newblock Klee: unassisted and automatic generation of high-coverage tests for complex systems programs.
\newblock In \emph{OSDI}, volume~8, pp.\  209--224, 2008.

\bibitem[Cha et~al.(2015)Cha, Woo, and Brumley]{cha2015program}
Sang~Kil Cha, Maverick Woo, and David Brumley.
\newblock Program-adaptive mutational fuzzing.
\newblock In \emph{2015 IEEE Symposium on Security and Privacy}, pp.\  725--741. IEEE, 2015.

\bibitem[Chae et~al.(2024)Chae, Kwon, Moon, Song, Kang, Ong, Kwak, Bae, Hwang, and Yeo]{chae2024coffee}
Hyungjoo Chae, Taeyoon Kwon, Seungjun Moon, Yongho Song, Dongjin Kang, Kai Tzu-iunn Ong, Beong-woo Kwak, Seonghyeon Bae, Seung-won Hwang, and Jinyoung Yeo.
\newblock Coffee-gym: An environment for evaluating and improving natural language feedback on erroneous code.
\newblock \emph{arXiv preprint arXiv:2409.19715}, 2024.

\bibitem[Chen et~al.(2023{\natexlab{a}})Chen, Zhang, Nguyen, Zan, Lin, Lou, and Chen]{chencodet}
Bei Chen, Fengji Zhang, Anh Nguyen, Daoguang Zan, Zeqi Lin, Jian-Guang Lou, and Weizhu Chen.
\newblock Codet: Code generation with generated tests.
\newblock In \emph{The Eleventh International Conference on Learning Representations}, 2023{\natexlab{a}}.

\bibitem[Chen et~al.(2021)Chen, Tworek, Jun, Yuan, Pinto, Kaplan, Edwards, Burda, Joseph, Brockman, et~al.]{chen2021evaluating}
Mark Chen, Jerry Tworek, Heewoo Jun, Qiming Yuan, Henrique Ponde De~Oliveira Pinto, Jared Kaplan, Harri Edwards, Yuri Burda, Nicholas Joseph, Greg Brockman, et~al.
\newblock Evaluating large language models trained on code.
\newblock \emph{arXiv preprint arXiv:2107.03374}, 2021.

\bibitem[Chen et~al.(2023{\natexlab{b}})Chen, Lin, Sch{\"a}rli, and Zhou]{chen2023teaching}
Xinyun Chen, Maxwell Lin, Nathanael Sch{\"a}rli, and Denny Zhou.
\newblock Teaching large language models to self-debug.
\newblock \emph{arXiv preprint arXiv:2304.05128}, 2023{\natexlab{b}}.

\bibitem[Dettmers et~al.(2023)Dettmers, Pagnoni, Holtzman, and Zettlemoyer]{dettmers2023qlora}
Tim Dettmers, Artidoro Pagnoni, Ari Holtzman, and Luke Zettlemoyer.
\newblock Qlora: Efficient finetuning of quantized llms.
\newblock \emph{Advances in neural information processing systems}, 36:\penalty0 10088--10115, 2023.

\bibitem[Dubey et~al.(2024)Dubey, Jauhri, Pandey, Kadian, Al-Dahle, Letman, Mathur, Schelten, Yang, Fan, et~al.]{dubey2024llama}
Abhimanyu Dubey, Abhinav Jauhri, Abhinav Pandey, Abhishek Kadian, Ahmad Al-Dahle, Aiesha Letman, Akhil Mathur, Alan Schelten, Amy Yang, Angela Fan, et~al.
\newblock The llama 3 herd of models.
\newblock \emph{arXiv preprint arXiv:2407.21783}, 2024.

\bibitem[Ficco et~al.(2011)Ficco, Pietrantuono, and Russo]{ficco2011bug}
Massimo Ficco, Roberto Pietrantuono, and Stefano Russo.
\newblock Bug localization in test-driven development.
\newblock \emph{Advances in Software Engineering}, 2011\penalty0 (1):\penalty0 492757, 2011.

\bibitem[Gemini et~al.(2023)Gemini, Anil, Borgeaud, Alayrac, Yu, Soricut, Schalkwyk, Dai, Hauth, Millican, et~al.]{team2023gemini}
Team Gemini, Rohan Anil, Sebastian Borgeaud, Jean-Baptiste Alayrac, Jiahui Yu, Radu Soricut, Johan Schalkwyk, Andrew~M Dai, Anja Hauth, Katie Millican, et~al.
\newblock Gemini: a family of highly capable multimodal models.
\newblock \emph{arXiv preprint arXiv:2312.11805}, 2023.

\bibitem[Gu et~al.(2024)Gu, Rozi{\`e}re, Leather, Solar-Lezama, Synnaeve, and Wang]{gu2024cruxeval}
Alex Gu, Baptiste Rozi{\`e}re, Hugh Leather, Armando Solar-Lezama, Gabriel Synnaeve, and Sida~I Wang.
\newblock Cruxeval: A benchmark for code reasoning, understanding and execution.
\newblock \emph{arXiv preprint arXiv:2401.03065}, 2024.

\bibitem[Guo et~al.(2024{\natexlab{a}})Guo, Zhu, Yang, Xie, Dong, Zhang, Chen, Bi, Wu, Li, Luo, Xiong, and Liang]{deepseek-coder}
Daya Guo, Qihao Zhu, Dejian Yang, Zhenda Xie, Kai Dong, Wentao Zhang, Guanting Chen, Xiao Bi, Y.~Wu, Y.K. Li, Fuli Luo, Yingfei Xiong, and Wenfeng Liang.
\newblock Deepseek-coder: When the large language model meets programming -- the rise of code intelligence, 2024{\natexlab{a}}.
\newblock URL \url{https://arxiv.org/abs/2401.14196}.

\bibitem[Guo et~al.(2024{\natexlab{b}})Guo, Zhu, Yang, Xie, Dong, Zhang, Chen, Bi, Wu, Li, et~al.]{guo2024deepseek}
Daya Guo, Qihao Zhu, Dejian Yang, Zhenda Xie, Kai Dong, Wentao Zhang, Guanting Chen, Xiao Bi, Yu~Wu, YK~Li, et~al.
\newblock Deepseek-coder: When the large language model meets programming--the rise of code intelligence.
\newblock \emph{arXiv preprint arXiv:2401.14196}, 2024{\natexlab{b}}.

\bibitem[Holler et~al.(2012)Holler, Herzig, and Zeller]{holler2012fuzzing}
Christian Holler, Kim Herzig, and Andreas Zeller.
\newblock Fuzzing with code fragments.
\newblock In \emph{21st USENIX Security Symposium (USENIX Security 12)}, pp.\  445--458, 2012.

\bibitem[Hu et~al.(2021)Hu, Shen, Wallis, Allen-Zhu, Li, Wang, Wang, and Chen]{hu2021lora}
Edward~J Hu, Yelong Shen, Phillip Wallis, Zeyuan Allen-Zhu, Yuanzhi Li, Shean Wang, Lu~Wang, and Weizhu Chen.
\newblock Lora: Low-rank adaptation of large language models.
\newblock \emph{arXiv preprint arXiv:2106.09685}, 2021.

\bibitem[Huang et~al.(2024)Huang, Chen, Mishra, Zheng, Yu, Song, and Zhou]{huang2024large}
Jie Huang, Xinyun Chen, Swaroop Mishra, Huaixiu~Steven Zheng, Adams~Wei Yu, Xinying Song, and Denny Zhou.
\newblock Large language models cannot self-correct reasoning yet.
\newblock In \emph{The Twelfth International Conference on Learning Representations}, 2024.
\newblock URL \url{https://openreview.net/forum?id=IkmD3fKBPQ}.

\bibitem[Hui et~al.(2024)Hui, Yang, Cui, Yang, Liu, Zhang, Liu, Zhang, Yu, Dang, et~al.]{hui2024qwen2}
Binyuan Hui, Jian Yang, Zeyu Cui, Jiaxi Yang, Dayiheng Liu, Lei Zhang, Tianyu Liu, Jiajun Zhang, Bowen Yu, Kai Dang, et~al.
\newblock Qwen2. 5-coder technical report.
\newblock \emph{arXiv preprint arXiv:2409.12186}, 2024.

\bibitem[Hurst et~al.(2024)Hurst, Lerer, Goucher, Perelman, Ramesh, Clark, Ostrow, Welihinda, Hayes, Radford, et~al.]{hurst2024gpt}
Aaron Hurst, Adam Lerer, Adam~P Goucher, Adam Perelman, Aditya Ramesh, Aidan Clark, AJ~Ostrow, Akila Welihinda, Alan Hayes, Alec Radford, et~al.
\newblock Gpt-4o system card.
\newblock \emph{arXiv preprint arXiv:2410.21276}, 2024.

\bibitem[Jain et~al.(2024)Jain, Han, Gu, Li, Yan, Zhang, Wang, Solar-Lezama, Sen, and Stoica]{jain2024livecodebench}
Naman Jain, King Han, Alex Gu, Wen-Ding Li, Fanjia Yan, Tianjun Zhang, Sida Wang, Armando Solar-Lezama, Koushik Sen, and Ion Stoica.
\newblock Livecodebench: Holistic and contamination free evaluation of large language models for code.
\newblock \emph{arXiv preprint arXiv:2403.07974}, 2024.

\bibitem[Jimenez et~al.(2024)Jimenez, Yang, Wettig, Yao, Pei, Press, and Narasimhan]{swe_bench}
Carlos~E Jimenez, John Yang, Alexander Wettig, Shunyu Yao, Kexin Pei, Ofir Press, and Karthik~R Narasimhan.
\newblock {SWE}-bench: Can language models resolve real-world github issues?
\newblock In \emph{The Twelfth International Conference on Learning Representations}, 2024.
\newblock URL \url{https://openreview.net/forum?id=VTF8yNQM66}.

\bibitem[King(1976)]{king1976symbolic}
James~C King.
\newblock Symbolic execution and program testing.
\newblock \emph{Communications of the ACM}, 19\penalty0 (7):\penalty0 385--394, 1976.

\bibitem[Lambert et~al.(2024{\natexlab{a}})Lambert, Morrison, Pyatkin, Huang, Ivison, Brahman, Miranda, Liu, Dziri, Lyu, Gu, Malik, Graf, Hwang, Yang, Bras, Tafjord, Wilhelm, Soldaini, Smith, Wang, Dasigi, and Hajishirzi]{lambert2024tulu3}
Nathan Lambert, Jacob Morrison, Valentina Pyatkin, Shengyi Huang, Hamish Ivison, Faeze Brahman, Lester James~V. Miranda, Alisa Liu, Nouha Dziri, Shane Lyu, Yuling Gu, Saumya Malik, Victoria Graf, Jena~D. Hwang, Jiangjiang Yang, Ronan~Le Bras, Oyvind Tafjord, Chris Wilhelm, Luca Soldaini, Noah~A. Smith, Yizhong Wang, Pradeep Dasigi, and Hannaneh Hajishirzi.
\newblock Tülu 3: Pushing frontiers in open language model post-training.
\newblock 2024{\natexlab{a}}.

\bibitem[Lambert et~al.(2024{\natexlab{b}})Lambert, Pyatkin, Morrison, Miranda, Lin, Chandu, Dziri, Kumar, Zick, Choi, Smith, and Hajishirzi]{RewardBench}
Nathan Lambert, Valentina Pyatkin, Jacob Morrison, LJ~Miranda, Bill~Yuchen Lin, Khyathi Chandu, Nouha Dziri, Sachin Kumar, Tom Zick, Yejin Choi, Noah~A. Smith, and Hannaneh Hajishirzi.
\newblock Rewardbench: Evaluating reward models for language modeling, 2024{\natexlab{b}}.
\newblock URL \url{https://huggingface.co/spaces/allenai/reward-bench}.

\bibitem[Li et~al.(2023)Li, Allal, Zi, Muennighoff, Kocetkov, Mou, Marone, Akiki, Li, Chim, et~al.]{li2023starcoder}
Raymond Li, Loubna~Ben Allal, Yangtian Zi, Niklas Muennighoff, Denis Kocetkov, Chenghao Mou, Marc Marone, Christopher Akiki, Jia Li, Jenny Chim, et~al.
\newblock Starcoder: may the source be with you!
\newblock \emph{arXiv preprint arXiv:2305.06161}, 2023.

\bibitem[Li et~al.(2022)Li, Choi, Chung, Kushman, Schrittwieser, Leblond, Eccles, Keeling, Gimeno, Dal~Lago, et~al.]{li2022competition}
Yujia Li, David Choi, Junyoung Chung, Nate Kushman, Julian Schrittwieser, R{\'e}mi Leblond, Tom Eccles, James Keeling, Felix Gimeno, Agustin Dal~Lago, et~al.
\newblock Competition-level code generation with alphacode.
\newblock \emph{Science}, 378\penalty0 (6624):\penalty0 1092--1097, 2022.

\bibitem[Lightman et~al.(2023)Lightman, Kosaraju, Burda, Edwards, Baker, Lee, Leike, Schulman, Sutskever, and Cobbe]{lightman2023let}
Hunter Lightman, Vineet Kosaraju, Yura Burda, Harri Edwards, Bowen Baker, Teddy Lee, Jan Leike, John Schulman, Ilya Sutskever, and Karl Cobbe.
\newblock Let's verify step by step.
\newblock \emph{arXiv preprint arXiv:2305.20050}, 2023.

\bibitem[Liu et~al.(2024{\natexlab{a}})Liu, Zeng, Liu, Yan, He, Wang, Yan, Liu, and Zhou]{liu2024skywork}
Chris~Yuhao Liu, Liang Zeng, Jiacai Liu, Rui Yan, Jujie He, Chaojie Wang, Shuicheng Yan, Yang Liu, and Yahui Zhou.
\newblock Skywork-reward: Bag of tricks for reward modeling in llms.
\newblock \emph{arXiv preprint arXiv:2410.18451}, 2024{\natexlab{a}}.

\bibitem[Liu et~al.(2024{\natexlab{b}})Liu, Xia, Wang, and Zhang]{liu2024your}
Jiawei Liu, Chunqiu~Steven Xia, Yuyao Wang, and Lingming Zhang.
\newblock Is your code generated by chatgpt really correct? rigorous evaluation of large language models for code generation.
\newblock \emph{Advances in Neural Information Processing Systems}, 36, 2024{\natexlab{b}}.

\bibitem[Madaan et~al.(2024)Madaan, Tandon, Gupta, Hallinan, Gao, Wiegreffe, Alon, Dziri, Prabhumoye, Yang, et~al.]{madaan2024self}
Aman Madaan, Niket Tandon, Prakhar Gupta, Skyler Hallinan, Luyu Gao, Sarah Wiegreffe, Uri Alon, Nouha Dziri, Shrimai Prabhumoye, Yiming Yang, et~al.
\newblock Self-refine: Iterative refinement with self-feedback.
\newblock \emph{Advances in Neural Information Processing Systems}, 36, 2024.

\bibitem[Mahan et~al.(2024)Mahan, Van~Phung, Rafailov, Blagden, Lile, Castricato, Fr{\"a}nken, Finn, and Albalak]{mahan2024generative}
Dakota Mahan, Duy Van~Phung, Rafael Rafailov, Chase Blagden, Nathan Lile, Louis Castricato, Jan-Philipp Fr{\"a}nken, Chelsea Finn, and Alon Albalak.
\newblock Generative reward models.
\newblock \emph{arXiv preprint arXiv:2410.12832}, 2024.

\bibitem[Maximilien \& Williams(2003)Maximilien and Williams]{tdd_at_ibm}
E.M. Maximilien and L.~Williams.
\newblock Assessing test-driven development at ibm.
\newblock In \emph{25th International Conference on Software Engineering, 2003. Proceedings.}, pp.\  564--569, 2003.
\newblock \doi{10.1109/ICSE.2003.1201238}.

\bibitem[Moon et~al.(2023)Moon, Chae, Song, Kwon, Kang, Ong, Hwang, and Yeo]{moon2023coffee}
Seungjun Moon, Hyungjoo Chae, Yongho Song, Taeyoon Kwon, Dongjin Kang, Kai Tzu-iunn Ong, Seung-won Hwang, and Jinyoung Yeo.
\newblock Coffee: Boost your code llms by fixing bugs with feedback.
\newblock \emph{arXiv preprint arXiv:2311.07215}, 2023.

\bibitem[Muennighoff et~al.(2023)Muennighoff, Liu, Zebaze, Zheng, Hui, Zhuo, Singh, Tang, von Werra, and Longpre]{octopack}
Niklas Muennighoff, Qian Liu, Armel Zebaze, Qinkai Zheng, Binyuan Hui, Terry~Yue Zhuo, Swayam Singh, Xiangru Tang, Leandro von Werra, and Shayne Longpre.
\newblock Octopack: Instruction tuning code large language models.
\newblock \emph{arXiv preprint arXiv:2308.07124}, 2023.

\bibitem[Muennighoff et~al.(2024)Muennighoff, Liu, Zebaze, Zheng, Hui, Zhuo, Singh, Tang, Werra, and Longpre]{muennighoff2024octopack}
Niklas Muennighoff, Qian Liu, Armel~Randy Zebaze, Qinkai Zheng, Binyuan Hui, Terry~Yue Zhuo, Swayam Singh, Xiangru Tang, Leandro~Von Werra, and Shayne Longpre.
\newblock Octopack: Instruction tuning code large language models.
\newblock In \emph{The Twelfth International Conference on Learning Representations}, 2024.
\newblock URL \url{https://openreview.net/forum?id=mw1PWNSWZP}.

\bibitem[Nagappan et~al.(2008)Nagappan, Maximilien, Bhat, and Williams]{quality_improvement_tdd}
Nachiappan Nagappan, E.~Michael Maximilien, Thirumalesh Bhat, and Laurie~A. Williams.
\newblock Realizing quality improvement through test driven development: results and experiences of four industrial teams.
\newblock \emph{Empir. Softw. Eng.}, 13\penalty0 (3):\penalty0 289--302, 2008.
\newblock \doi{10.1007/S10664-008-9062-Z}.
\newblock URL \url{https://doi.org/10.1007/s10664-008-9062-z}.

\bibitem[Ni et~al.(2024)Ni, Allamanis, Cohan, Deng, Shi, Sutton, and Yin]{ninext}
Ansong Ni, Miltiadis Allamanis, Arman Cohan, Yinlin Deng, Kensen Shi, Charles Sutton, and Pengcheng Yin.
\newblock Next: Teaching large language models to reason about code execution.
\newblock In \emph{Forty-first International Conference on Machine Learning}, 2024.

\bibitem[Olausson et~al.(2023)Olausson, Inala, Wang, Gao, and Solar-Lezama]{olausson2023self}
Theo~X Olausson, Jeevana~Priya Inala, Chenglong Wang, Jianfeng Gao, and Armando Solar-Lezama.
\newblock Is self-repair a silver bullet for code generation?
\newblock In \emph{The Twelfth International Conference on Learning Representations}, 2023.

\bibitem[Prasad et~al.(2024)Prasad, Yuan, Pang, Xu, Fazel-Zarandi, Bansal, Sukhbaatar, Weston, and Yu]{prasad2024self}
Archiki Prasad, Weizhe Yuan, Richard~Yuanzhe Pang, Jing Xu, Maryam Fazel-Zarandi, Mohit Bansal, Sainbayar Sukhbaatar, Jason Weston, and Jane Yu.
\newblock Self-consistency preference optimization.
\newblock \emph{arXiv preprint arXiv:2411.04109}, 2024.

\bibitem[Rice(1953)]{rices_theorem}
Henry~Gordon Rice.
\newblock Classes of recursively enumerable sets and their decision problems.
\newblock \emph{Transactions of the American Mathematical Society}, 74\penalty0 (2):\penalty0 358--366, 1953.

\bibitem[Roziere et~al.(2023)Roziere, Gehring, Gloeckle, Sootla, Gat, Tan, Adi, Liu, Sauvestre, Remez, et~al.]{roziere2023code}
Baptiste Roziere, Jonas Gehring, Fabian Gloeckle, Sten Sootla, Itai Gat, Xiaoqing~Ellen Tan, Yossi Adi, Jingyu Liu, Romain Sauvestre, Tal Remez, et~al.
\newblock Code llama: Open foundation models for code.
\newblock \emph{arXiv preprint arXiv:2308.12950}, 2023.

\bibitem[Sch{\"a}fer et~al.(2023)Sch{\"a}fer, Nadi, Eghbali, and Tip]{schafer2023empirical}
Max Sch{\"a}fer, Sarah Nadi, Aryaz Eghbali, and Frank Tip.
\newblock An empirical evaluation of using large language models for automated unit test generation.
\newblock \emph{IEEE Transactions on Software Engineering}, 2023.

\bibitem[Shinn et~al.(2024)Shinn, Cassano, Gopinath, Narasimhan, and Yao]{shinn2024reflexion}
Noah Shinn, Federico Cassano, Ashwin Gopinath, Karthik Narasimhan, and Shunyu Yao.
\newblock Reflexion: Language agents with verbal reinforcement learning.
\newblock \emph{Advances in Neural Information Processing Systems}, 36, 2024.

\bibitem[Snell et~al.(2024)Snell, Lee, Xu, and Kumar]{snell2024scaling}
Charlie Snell, Jaehoon Lee, Kelvin Xu, and Aviral Kumar.
\newblock Scaling llm test-time compute optimally can be more effective than scaling model parameters.
\newblock \emph{arXiv preprint arXiv:2408.03314}, 2024.
\newblock URL \url{https://arxiv.org/abs/2408.03314}.

\bibitem[Team et~al.(2024)Team, Mesnard, Hardin, Dadashi, Bhupatiraju, Pathak, Sifre, Rivi{\`e}re, Kale, Love, et~al.]{team2024gemma}
Gemma Team, Thomas Mesnard, Cassidy Hardin, Robert Dadashi, Surya Bhupatiraju, Shreya Pathak, Laurent Sifre, Morgane Rivi{\`e}re, Mihir~Sanjay Kale, Juliette Love, et~al.
\newblock Gemma: Open models based on gemini research and technology.
\newblock \emph{arXiv preprint arXiv:2403.08295}, 2024.

\bibitem[Tillmann et~al.(2010)Tillmann, de~Halleux, and Xie]{parameterized_unit_testing}
Nikolai Tillmann, Jonathan de~Halleux, and Tao Xie.
\newblock Parameterized unit testing: theory and practice.
\newblock In \emph{2010 ACM/IEEE 32nd International Conference on Software Engineering}, volume~2, pp.\  483--484, 2010.
\newblock \doi{10.1145/1810295.1810441}.

\bibitem[Wang et~al.(2022)Wang, Wei, Schuurmans, Le, Chi, Narang, Chowdhery, and Zhou]{wang2022self}
Xuezhi Wang, Jason Wei, Dale Schuurmans, Quoc~V Le, Ed~H Chi, Sharan Narang, Aakanksha Chowdhery, and Denny Zhou.
\newblock Self-consistency improves chain of thought reasoning in language models.
\newblock In \emph{The Eleventh International Conference on Learning Representations}, 2022.

\bibitem[Wei et~al.(2022)Wei, Wang, Schuurmans, Bosma, Xia, Chi, Le, Zhou, et~al.]{wei2022chain}
Jason Wei, Xuezhi Wang, Dale Schuurmans, Maarten Bosma, Fei Xia, Ed~Chi, Quoc~V Le, Denny Zhou, et~al.
\newblock Chain-of-thought prompting elicits reasoning in large language models.
\newblock \emph{Advances in neural information processing systems}, 35:\penalty0 24824--24837, 2022.

\bibitem[Zelikman et~al.(2022)Zelikman, Wu, Mu, and Goodman]{zelikman2022star}
Eric Zelikman, Yuhuai Wu, Jesse Mu, and Noah Goodman.
\newblock {STaR}: Bootstrapping reasoning with reasoning.
\newblock \emph{Advances in Neural Information Processing Systems}, 35:\penalty0 15476--15488, 2022.

\bibitem[Zhang et~al.(2023)Zhang, Li, Li, Li, and Jin]{zhang2023self}
Kechi Zhang, Zhuo Li, Jia Li, Ge~Li, and Zhi Jin.
\newblock Self-edit: Fault-aware code editor for code generation.
\newblock \emph{arXiv preprint arXiv:2305.04087}, 2023.

\bibitem[Zhang et~al.(2024)Zhang, Hosseini, Bansal, Kazemi, Kumar, and Agarwal]{zhang2024generative}
Lunjun Zhang, Arian Hosseini, Hritik Bansal, Mehran Kazemi, Aviral Kumar, and Rishabh Agarwal.
\newblock Generative verifiers: Reward modeling as next-token prediction.
\newblock \emph{arXiv preprint arXiv:2408.15240}, 2024.

\bibitem[Zhong et~al.(2024)Zhong, Wang, and Shang]{zhong2024ldb}
Li~Zhong, Zilong Wang, and Jingbo Shang.
\newblock Ldb: A large language model debugger via verifying runtime execution step-by-step.
\newblock \emph{arXiv preprint arXiv:2402.16906}, 2024.

\end{thebibliography}
